\documentclass[iop]{emulateapj}
\usepackage{natbib}
\usepackage{graphics}

\usepackage{lscape}

\shorttitle{Isochrones and IL Spectroscopy}
\shortauthors{COLUCCI \& BERNSTEIN}

\begin{document}
\newcommand{\msol}{M_\odot}
\newcommand{\etal}{et al.\ }
\newcommand{\kms}{km~s$^{-1}$ }
\newcommand{\rAA}{{\AA \enskip}}
\newcommand{\ew}{W_\lambda}
\newcommand{\mv}{$M_{V}^{\rm{tot}}$}
\newcommand{\bvo}{($B-V$)$_{\rm{0}}$} 
\defcitealias{mb08}{MB08}
\defcitealias{milkyway}{Paper II}
\defcitealias{paper3}{C11}
\newcommand{\rkms}{km~s$^{-1}$ \enskip}

\title{Comparison of Convective Overshooting Models and Their Impact on Abundances  from Integrated Light Spectroscopy of Young ($<$ 3 Gyr) Star Clusters\footnotemark[1]}

\footnotetext[1]{This paper includes data gathered with the 6.5 meter Magellan 
Telescopes located at Las Campanas Observatory, Chile.}

\author{Janet E. Colucci}
\affil{Department of Astronomy and Astrophysics, 1156  High Street, UCO/Lick Observatory, \\ University of California, Santa Cruz, CA 95064; jcolucci@ucolick.org}

\and

\author{Rebecca A. Bernstein}
\affil{Department of Astronomy and Astrophysics, 1156  High Street, UCO/Lick Observatory,\\
 University of California, Santa Cruz, CA 95064; rab@ucolick.org}

\begin{abstract}
As part of an ongoing program to measure detailed chemical abundances in nearby galaxies, we use a sample of young to intermediate age clusters in the Large Magellanic Cloud with ages of 10 Myr  to 2 Gyr to evaluate the effect of isochrone parameters, specifically core convective overshooting,  on Fe abundance results from  high resolution, integrated light spectroscopy.   In this work we also obtain fiducial Fe abundances from  high resolution spectroscopy of the cluster individual member stars.   We compare the Fe abundance results for the individual stars to the results from isochrones and integrated light spectroscopy to determine whether isochrones with convective overshooting should be used in our  integrated light analysis of young  to intermediate age (10 Myr -3 Gyr) star clusters.  We find that when using the isochrones from the Teramo group, we obtain  more accurate results for young and intermediate age clusters over the entire age range  when using isochrones without convective overshooting.  While convective overshooting is not the only uncertain aspect of stellar evolution, it is one of the most readily parametrized ingredients in  stellar evolution models, and thus important to evaluate for the specific models used in our integrated light analysis.   This work demonstrates that our method for integrated light spectroscopy of star clusters can provide  unique  tests for future constraints on stellar evolution models of young and intermediate age clusters.

\end{abstract}


\section{Introduction}
\label{sec:intro}
\setcounter{footnote}{1}

 We  are conducting 
 a study of Large Magellanic Cloud (LMC) clusters in the young to  intermediate age range with the ultimate goal of obtaining detailed abundances of over 20 elements from integrated light (IL), high resolution spectroscopy. Our method uses stellar evolution models and  isochrones to create representative color magnitude diagrams (CMDs), with which to execute detailed spectral synthesis.		
To understand our accuracy we must evaluate to what extent  uncertainties in stellar evolution modeling can affect our results.
 In \cite{m31paper} and \citet{paper3} (hereafter \citetalias{paper3}),  we evaluated relevant issues for old ($>$5 Gyr) clusters, mainly uncertainties in  horizontal branch morphology and asymptotic giant branch stars.   In this work, we address additional challenges in analysis of young and intermediate age clusters  (age $<$3 Gyr), concentrating on  the effects of    core convective overshooting  in the isochrones.  While the inclusion of convective overshooting is not the only uncertainty in stellar models for young stars \citep[for example, see][]{ventura05}, it is a key parameter causing significant differences in evolution of young stars; consequently stellar modeling groups do  tabulate isochrone families with different values of convective overshooting. 

 Briefly, convective overshooting (C-OVER) refers to the treatment of convection at the border of stellar cores.
Stars that are more massive than $\sim$1.1 $\msol$ are hot enough to develop a convective core during the main sequence, H-burning phase.   Note that, because only relatively massive stars develop convective cores, the treatment of C-OVER is only important for star clusters that are younger than  $\sim$3 Gyr. Historically,  two types of treatments have been used in stellar evolution models to describe stars with convective stellar cores.
The first, most simplistic model, uses the Schwarzschild criterion to treat convective instability.  In this case there is a clean boundary at the edge of the convective stellar core, and the stellar properties (luminosity, lifetime, etc.) are determined by the input stellar physics.The second treatment parametrizes a certain amount of mechanical convective overshooting  past the Schwarzschild core boundary.
 The addition of C-OVER into the stellar evolution models is physically motivated by the fact that fluid elements could maintain some velocity when moving past the Schwarzschild boundary,  due to residual momentum or the star's rotational velocity
  \citep[e.g.][]{2000A&A...361..101M}. In  stellar evolution models, the empirical effect of including C-OVER is an increase in the size of the stellar core, which results in a higher stellar luminosity and a shorter stellar lifetime.

\begin{figure*}
\centering
\includegraphics[angle=90,scale=0.6]{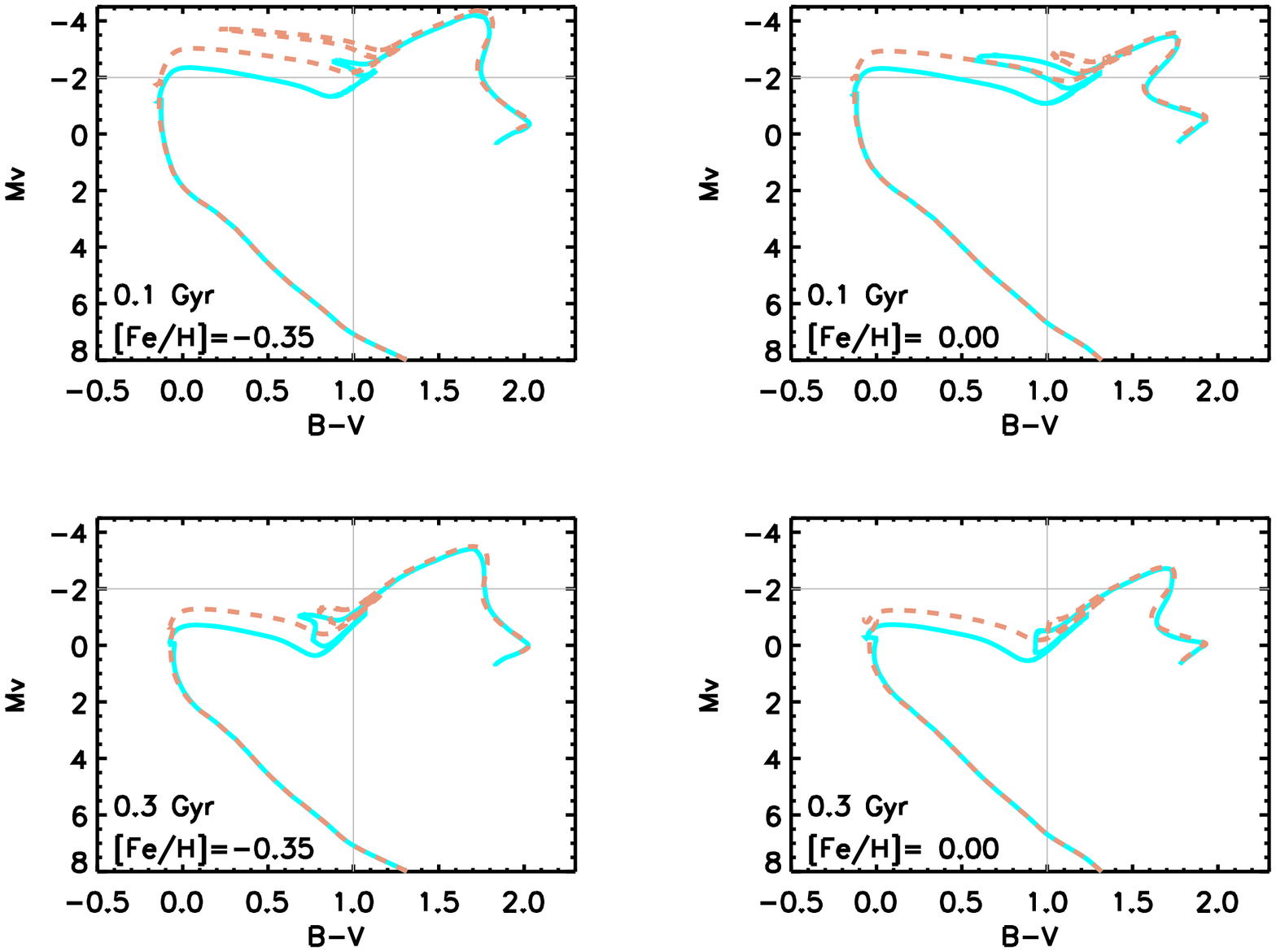}
\includegraphics[angle=90,scale=0.6]{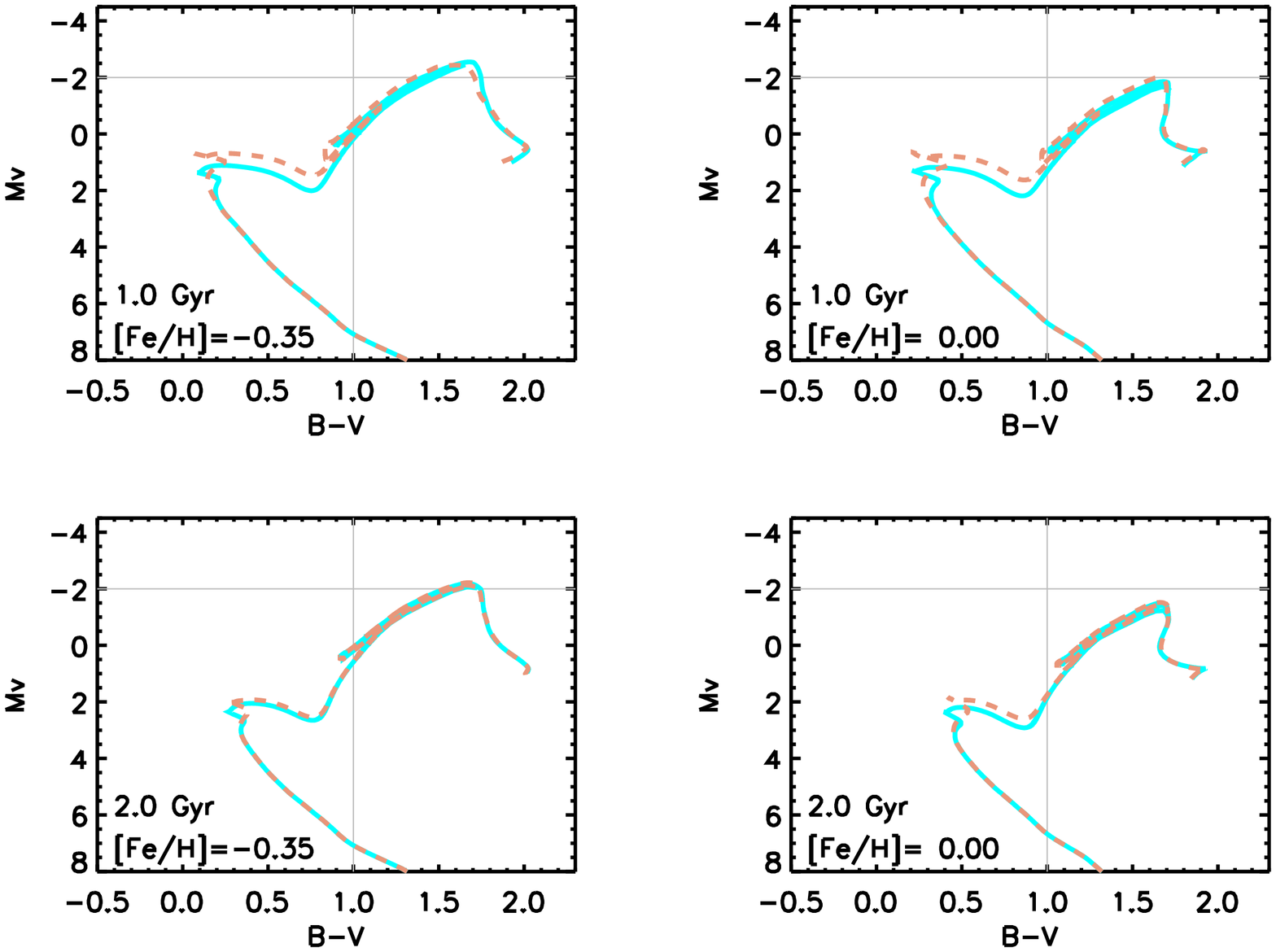}
\caption{Comparison of Teramo isochrones with no C-OVER (cyan) and with C-OVER (pink) for ages of 0.1 to 2 Gyr and metallicities of [Fe/H]=$-0.35$ and [Fe/H]=$0.00$, as shown. Gray lines are shown to guide the eye.  For ages $<$1 Gyr the inclusion of C-OVER results in large differences in the supergiant populations, while for ages $>$1 Gyr there are only subtle differences in the turnoff morphology.  The differences between the models are typically larger at lower metallicity.}
\label{fig:isos} 
\end{figure*}

 Observations of stellar clusters have been used to try to constrain the appropriateness and magnitude of  C-OVER that should be included in stellar evolution models.	 Because including C-OVER in the stellar models increases the luminosities of  supergiant stars, which have ages of $\sim$100's of Myrs, C-OVER models in this age range will predict older ages  for stellar clusters  than models without C-OVER.  This is visually apparent in Figure \ref{fig:isos}, where we show examples of Teramo  \citep{2004ApJ...612..168P}  isochrones with and without C-OVER for ages of 0.1, 0.3, 1.0, and 2.0 Gyr, and metallicities of [Fe/H]$=0$ and [Fe/H]$=-0.35$.  For ages of 0.1 and 0.3 Gyr, the turnoff stars and blue loop supergiants have brighter magnitudes when C-OVER is included  (shown by the pink dashed lines).   Figure \ref{fig:isos} also demonstrates that the differences in the giant populations of the different sets of isochrones become smaller as the age of the isochrones increases. For ages of $\sim$2 Gyr, the giant populations are very similar;   only a small difference in the turnoff morphologies is evident.
    Finally, Figure \ref{fig:isos} shows that, with the exception of the 2 Gyr case,  the differences between isochrones with and without C-OVER are greater for the lower metallicity of [Fe/H]=$-0.35$, which is roughly the present day metallicity of the LMC, than they are for higher metallicities of [Fe/H]=$0$.

While  C-OVER was introduced into stellar evolution models to reproduce observations  of stars in young stellar clusters, the  appropriateness and magnitude of C-OVER that is required  is still under debate.
 For ages of $\sim$150 Myr, where differences are most dramatic for supergiant stars, various authors have tried to use the CMD of NGC 1866 to constrain the physics of the stellar evolution models. Different authors have reached different conclusions, and the results appear to be dependent on the set of stellar models that are used.  For example,  using a ground based CMD and FRANEC \citep{franec} models, 			 \cite{1999AJ....118.2839T} determined that C-OVER was not required to match  observations of NGC 1866.
 On the other hand, \cite{2002A&A...385..847B}, using the same dataset as  \cite{1999AJ....118.2839T}, found that when using  the Padova \citep{padova} models,  it is necessary to include  C-OVER in order to reproduce the observations.
Later,  \cite{2003AJ....125.3111B} and \cite{2004MmSAI..75..142B} used a Hubble Space Telescope (HST) WFPC2 CMD of NGC 1866 to  determine the best fitting  degree of C-OVER with higher quality data. 
They found that C-OVER was not required to fit the observations using the Pisa \citep{pisa} stellar models, but that when Padova models are used, including C-OVER still provides a better fit, as found using the ground based CMDs.
Unfortunately, to the best of our knowledge, the Teramo \citep{2004ApJ...612..168P} models with and without C-OVER that we use in our IL analysis have not been compared to observations of NGC 1866.

For clusters with older ages ($\sim$ 2 Gyrs),  high precision HST ACS photometry was analyzed by  \cite{2007AJ....133.2053M} and  \cite{2007AJ....134.1813M}  for the LMC clusters NGC 1978 and NGC 1783.  These authors tested stellar evolution models from several different groups, including the Teramo isochrones, and concentrated on the turnoff regions of the clusters, where differences are most visible in this age range.     After testing the Teramo, Pisa, and Padova models, the authors found that some amount of C-OVER is required to match the turnoff region, regardless of which set of models were used.

\subsection{Background:  The Impact on High Resolution Integrated Light Spectroscopy}
\label{sec:isos}

	 As described above, we use a grid of  isochrones when measuring abundances with   our high resolution,  integrated light spectra abundance analysis method \citep[see][ for full details]{bernstein02,mb08,scottphd,mythesis,m31paper,paper3}.
 For long-term consistency, it is crucial that we  choose  a single group of isochrones to use in all of our analyses.  We have selected those of the Teramo group \citep{2004ApJ...612..168P,2006ApJ...642..797P}, because  they provide   a wide range of ages, metallicities and two levels of $\alpha$-enhancement. The Teramo isochrones are also tabulated with two values of C-OVER:   no C-OVER (listed as  canonical) and with moderate C-OVER (listed as  non-canonical).

 Our initial work concerned typical old globular clusters with ages $>$5 Gyr, which are not sensitive to inclusion of C-OVER because the stars in old clusters are not massive enough to have convective cores. However, in \citetalias{paper3} we extended our method to young clusters in the LMC,   and it became clear that our results for the youngest clusters (ages $<$1 Gyr) could be significantly affected by the choice of C-OVER.
In \citetalias{paper3},  we chose to  adopt the Teramo isochrones with C-OVER included, because \cite{2007AJ....133.2053M} determined that these isochrones were a better match to the CMD of NGC 1978, a cluster that is also in our sample, than isochrones without C-OVER.  Unfortunately, there was no consensus in the literature for the use of C-OVER for clusters over the entire age range of the young clusters in our sample (10 Myr to 2 Gyr).  While several authors concluded that no C-OVER was needed in order to match observations of NGC 1866 (age 150 Myr), none of these authors tested isochrones from the Teramo group.  Note that this result is at odds with the results for NGC 1978 (age 2 Gyr).   We mentioned above that because C-OVER is not the only uncertainty in stellar evolution physics, conclusions regarding the need for C-OVER are dependent on the set of models being used.  Therefore, in order  for us to determine the ``right'' set of models to use in  our new IL analysis method {\it over the entire cluster age range} we have to determine which set of models produces abundances for our sample of local clusters  that are closest to the abundances for these clusters determined by other well-established methods.

In \citetalias{paper3}, initial tests of isochrones with and without C-OVER for clusters with ages of 2 Gyr showed that the differences were smaller than the statistical uncertainties of the solutions. However, results for clusters with ages $<$1 Gyr were not always in agreement with the few results available in the literature.  Unfortunately, more than half of the young clusters in our sample had no previous abundance measurements from high resolution abundance analysis of  individual stars.	Moreover, because individual star analyses performed by different authors do not agree to better than $\sim0.10-0.15$ dex \citep{2004ARA&A..42..385G,scottphd}, we cannot be certain which differences can be attributed to isochrone physics, and which are due to systematic offsets between different abundance analysis techniques.   We have therefore completed our own analysis of individual stars in the young clusters in our sample, to obtain abundances using identical  line lists, stellar atmospheres, and spectral synthesis codes as in our IL spectra analysis.    We identify the most accurate set of isochrones as those that produce the Fe abundances closest to the Fe abundances that we measure for the individual stars analyzed in this work.

We note that in the literature convective overshooting has been extensively studied using a much  larger sample of clusters than we describe here, including many nearby open clusters in the Milky Way.  We emphasize that our principle intent is to specifically evaluate the most appropriate set of Teramo isochrones for use in our high resolution integrated light spectra technique.  In this regard, we refrain from any analysis or comparison of Teramo isochrones to  resolved photometric color magnitude diagrams of the LMC clusters in our sample.  While such an analysis would be very interesting to compare to the results here, it is beyond the scope of the  present spectroscopic analysis and does not address our primary goal of determining which Teramo isochrones result in integrated light Fe abundances that most closely match fiducial Fe abundances from individual stars.

This paper is organized as follows: in \textsection\ref{sec:data} we describe the stellar and cluster targets and data reduction techniques. In \textsection\ref{sec:analysis} we present the  high resolution Fe abundance analysis for the individual stars in the LMC clusters, and review the analysis for the integrated light spectra of the clusters originally presented in  \citetalias{paper3}.    In \textsection \ref{sec:compare} we present the results,  discuss some of the sources of the differences in results when using different values of C-OVER in the IL analysis, and identify the set of isochrones for which the most accurate Fe abundances are obtained.

\section{Observations and Data Reduction}
\label{sec:data}

\subsection{Cluster Integrated Light}
\label{sec:data-clusters}

Our  integrated light spectra of NGC 1978, NGC 1866, NGC 1711, and NGC 2100 were obtained using the echelle spectrograph on the 2.5 m du Pont telescope at Las Campanas during  dark time in 2000 December and 2001 January. The wavelength coverage is approximately 3700--7800 \AA.  

 Integrated light spectra were obtained by  scanning a $12\times12$ arcsec$^{2}$ or  $8\times8$ arcsec$^{2}$ region of each cluster core \citep{mb08}.     These spectra were reduced with standard IRAF\footnote{IRAF is distributed
  by the National Optical Astronomy Observatories, which are operated
  by the Association of Universities for Research in Astronomy, Inc.,
  under cooperative agreement with the National Science
  Foundation.} routines, combined with the scattered-light subtraction described in \citet{mb08}. Complete details on the cluster integrated light observations and reductions can be found in \citet{mb08} and \citetalias{paper3}.

\subsection{Cluster Stars}
\label{sec:data-stars}

Stars were selected from the catalogs of  \cite{1995A&AS..112..367W},  \cite{1989ApJS...71...25B},    \cite{1991A&AS...90..387S}, and  \cite{1974A&AS...15..261R} for NGC 1978, NGC 1866, NGC 1711, and NGC 2100, respectively.  Information on the target stars, exposure times, and approximate signal-to-noise (S/N) ratios are given in Table \ref{tab:stellar_info}.

The spectra of individual stars in the LMC clusters were obtained with the MIKE double echelle spectrograph \citep{mike} on the Magellan Clay Telescope during three different observing runs in 2003 and 2004.  The setup of the spectrograph changed  between the runs, which resulted in different wavelength coverages for the individual runs.  However, we primarily use lines with wavelengths between 4500-7500 \rAA (red side only) in our analysis, which is a region in common to all three runs. 
The data taken in 2003 January used a 0.7''$\times$5'' slit and 4$\times$2 on chip binning.   The data taken in 2003 November used a  0.5"$\times$5'' slit and 3$\times$1 on chip binning, while the data from 2004 October used a  0.7''$\times$5'' slit and 3$\times$2 on chip binning.
The stellar spectra were reduced using the MIKE Redux\footnote{ http://www.ucolick.org/~\~xavier/IDL/index.html} pipeline \citep{mikeredux}, which includes a heliocentric velocity correction.  
 
Radial velocities of the stars are measured with the analysis code  GETJOB (see \textsection \ref{sec:ews}).  The radial velocities ($v_{r}$) are  calculated by determining velocity offsets in the spectra from a list of input strong lines.    Our averaged values are in good agreement with values in the literature.  For NGC 1978 we measure $v_{r}=291.5 \pm 2.0$ \kms, which agrees with the values of \cite{2000A&A...364L..19H},  \cite{2008AJ....136..375M}, and \cite{1991AJ....101..515O} of  $293.5 \pm 1.8$ $293.1 \pm 1.5$, and  $292.4 \pm 0.4$ \kms,  respectively.  For NGC 1866 we measure $v_{r}=301.5 \pm 1.1$ \kms, which agrees well with the values of \cite{2000A&A...364L..19H} and \cite{mucc1866} who found $299.8 \pm 1.4$ and $298.5 \pm 1.5$ \kms.  We measure  $v_{r}=244.2 \pm 2.8$  \rkms for NGC 1711, which is within 2 $\sigma$  of the value measured by \cite{1983ApJ...272..488F} of $230 \pm 9$ \kms. Similarly, our measurement of $v_{r}=247.6 \pm 5.7$ \rkms for NGC 2100 is within 2 $\sigma$ of the value of \cite{1994A&A...282..717J} of $262.5 \pm 6.45$ \kms.

\section{Abundance Analysis}
\label{sec:analysis}

Where possible, we have used identical analysis techniques for both the IL analysis of clusters and for the individual member stars.  In \textsection \ref{sec:ews} we describe the line lists and equivalent width measurements that both analyses have in common.  In \textsection \ref{sec:clusters} we briefly review the IL analysis of \citetalias{paper3}, and in \textsection \ref{sec:stars} we describe the new analysis of individual stars performed in this work.

\subsection{EWs and Line Lists}
\label{sec:ews}

As in our previous work \citepalias[e.g.][ and references therein]{paper3}, we use the semi-automated program GETJOB \citep{1995AJ....109.2736M} to measure absorption line equivalent widths (EWs)  for individual lines in both the cluster  IL  and stellar spectra in a consistent way.  Low order polynomials are  interactively fit to continuum regions for each spectral order, and line profiles are fit with single, double, or triple Gaussians as needed, depending on the presence of small line blends.  Line lists and log $gf$ values  are taken from \citetalias{paper3} and references therein. We only analyze Fe lines with EWs$<$150 m\rAA in order to minimize  line saturation effects.  Fe abundances are calculated under the assumption of local thermodynamic equilibrium (LTE).  The lines and EWs measured in the individual stars  are listed in  Table~\ref{tab:linetable_stars}.

\subsection{Cluster Integrated Light}
\label{sec:clusters}

In \citetalias{paper3} we presented  a detailed analysis of the integrated light  [Fe/H] and age solutions for
 each LMC cluster in our sample using Teramo isochrones with C-OVER included. In this work, we perform an identical analysis  but use the Teramo isochrones without C-OVER, as discussed in \textsection \ref{sec:intro}.   We summarize this analysis below.

 We create synthetic CMDs for the available range of age and metallicity of the Teramo isochrones, and   divide each CMD into $\sim$25 equal flux boxes containing stars of similar properties.   The properties of a flux-weighted  ``average'' star for each box are used in the IL EW synthesis, which we perform with ILABUNDS \citep{mb08}. ILABUNDs employs the 2010 version of  the spectral synthesis code MOOG \citep{moog}.  We use the ODFNEW  model stellar atmospheres of Kurucz%
\footnote{The models are available  from R. L. Kurucz's Website at  http://kurucz.harvard.edu/grids.html}
  \citep[e.g.][]{2004astro.ph..5087C} for all abundance analysis. We  choose the ODFNEW atmospheres instead of the AODFNEW atmospheres because we have determined that the clusters are not significantly $\alpha$-enhanced, as reported in \cite{paper4}.
  
To begin the analysis of any cluster, we calculate a mean [Fe/H]  abundance from all available Fe I  lines for the  large  grid of synthetic CMDs.  We note that because we  measure far fewer Fe II lines  than Fe I lines in the cluster IL spectra, and because of uncertainties in the relationship between neutral and ionized solutions,  we do not use Fe II lines to constrain the best-fitting CMD.  We  next use the quality of the Fe I abundance solution to constrain the best-fitting age and abundance for each cluster. Specifically, we determine  the best age and abundance using Fe line diagnostics \citep{mb08,m31paper,paper3}.  These diagnostics, also used in the standard stellar abundance analysis below,  relate to the quality of the [Fe/H] solutions.  In particular, a stable [Fe/H] solution should not depend on the parameters of individual lines (excitation potentials, wavelengths, or reduced EWs\footnote{Reduced EW 
$\equiv$ log(EW / $\lambda$) }), and the  standard deviation of the [Fe/H] solution  should be small.

 Finally,  we can improve our  solutions by  allowing for statistically incomplete sampling of the cluster CMDs \citepalias{paper3}. 
It is especially important to allow for statistical variations for clusters with ages under $\sim$2 Gyrs, because   clusters with these ages are  rapidly evolving,  and  in the case of our sample, they are   less luminous and  less massive  than typical Milky Way GCs,   and therefore the  most likely to suffer from stochastic effects. In order to quantify this uncertainty  we use a Monte Carlo technique to statistically populate the cluster initial mass functions (IMFs) with discrete numbers of stars, resulting in many possible realizations of each cluster.  As demonstrated in \citetalias{paper3}, the best statistical  realizations of each cluster can be identified using the Fe line diagnostics.  

Here, we briefly summarize the effect that the Monte Carlo tests have on the derived [Fe/H] and age of each cluster.   For NGC 1718 the [Fe/H] decreases marginally from -0.67 to -0.70, and the uncertainty due to the age solution is unchanged, at 0.03.  For NGC 1978,  we find that allowing for stochastic sampling decreases the derived [Fe/H] from -0.48 to -0.54, and increases the uncertainty due to the age from  0.05 to 0.18.  For NGC 1866, the [Fe/H] abundance is unchanged, but the uncertainty due to the age increases from 0.16 to 0.20.  For NGC 1711, both the derived [Fe/H] and uncertainty due to the age are unchanged.  As in \citetalias{paper3}, we are only able to derive a solution for NGC 2100 by allowing for statistical fluctuations, so there is no added uncertainty.

In summary,  the most self-consistent age and [Fe/H] solutions that we have determined for each cluster are listed in Table \ref{tab:abund_compare}.  In columns 2 and 3, we show the best-fitting age and  [Fe/H] solutions determined for isochrones with C-OVER in  \citetalias{paper3} and in columns 4 and 5 we show the best-fitting age and [Fe/H] solutions determined using isochrones without C-OVER from  this work.  We  note that we include results for the IL abundances of NGC 1718 in  Table \ref{tab:abund_compare}, but we do not have a sample of individual stars in this cluster and so the IL results are not used for the IL and stellar direct  comparison in \textsection \ref{sec:compare}. As already discussed, the measurement of the ages and metallicities from the IL spectra was explained in detail in \citetalias{paper3}, and is accompanied by an in depth comparison of the derived ages and [Fe/H] to values in the literature. We note that  the ages derived in this work using isochrones without C-OVER are consistent with the conclusions of \citetalias{paper3}, so we do not repeat that discussion here.

\begin{figure}
\centering
\includegraphics[angle=90,scale=0.3]{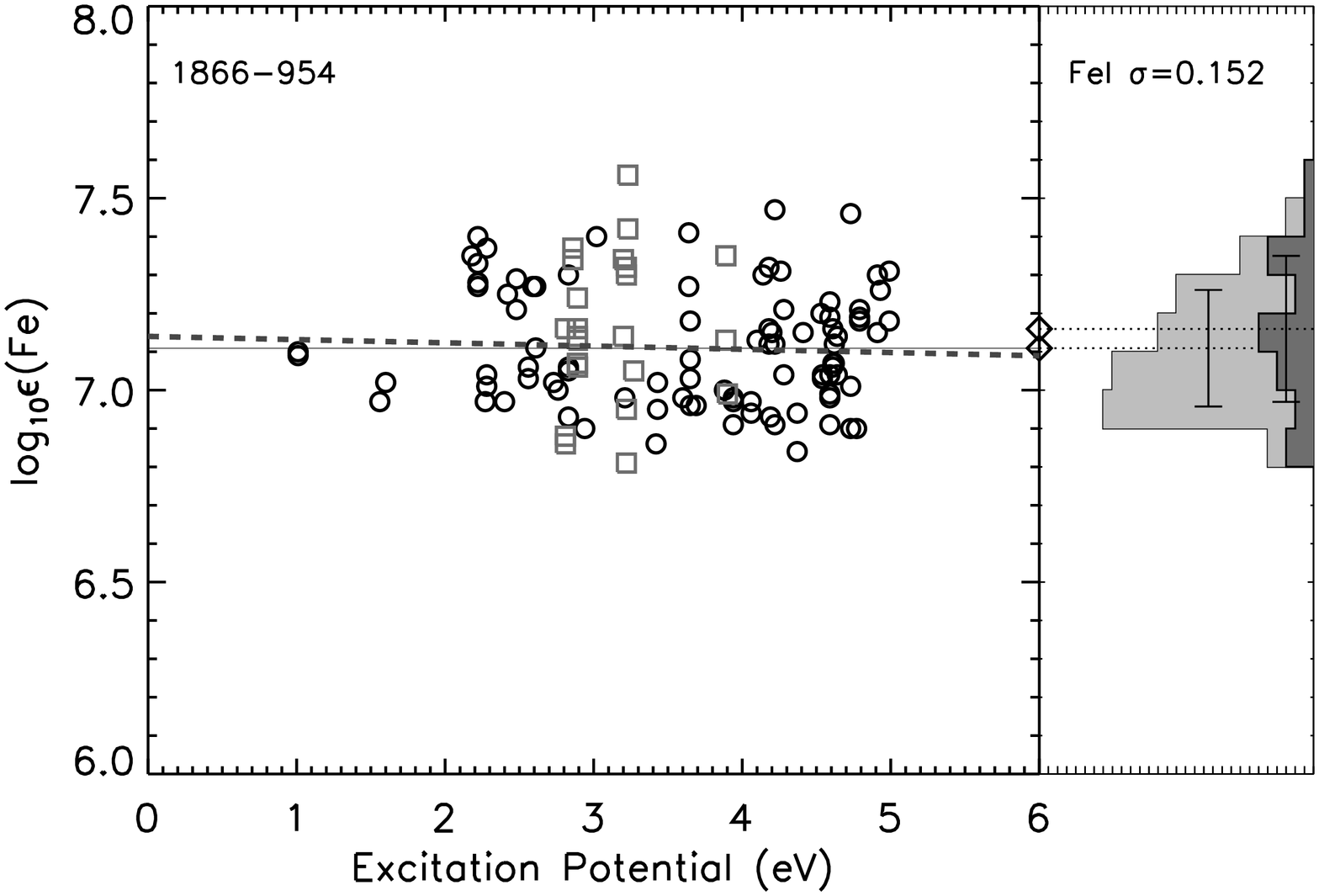}
\includegraphics[angle=90,scale=0.3]{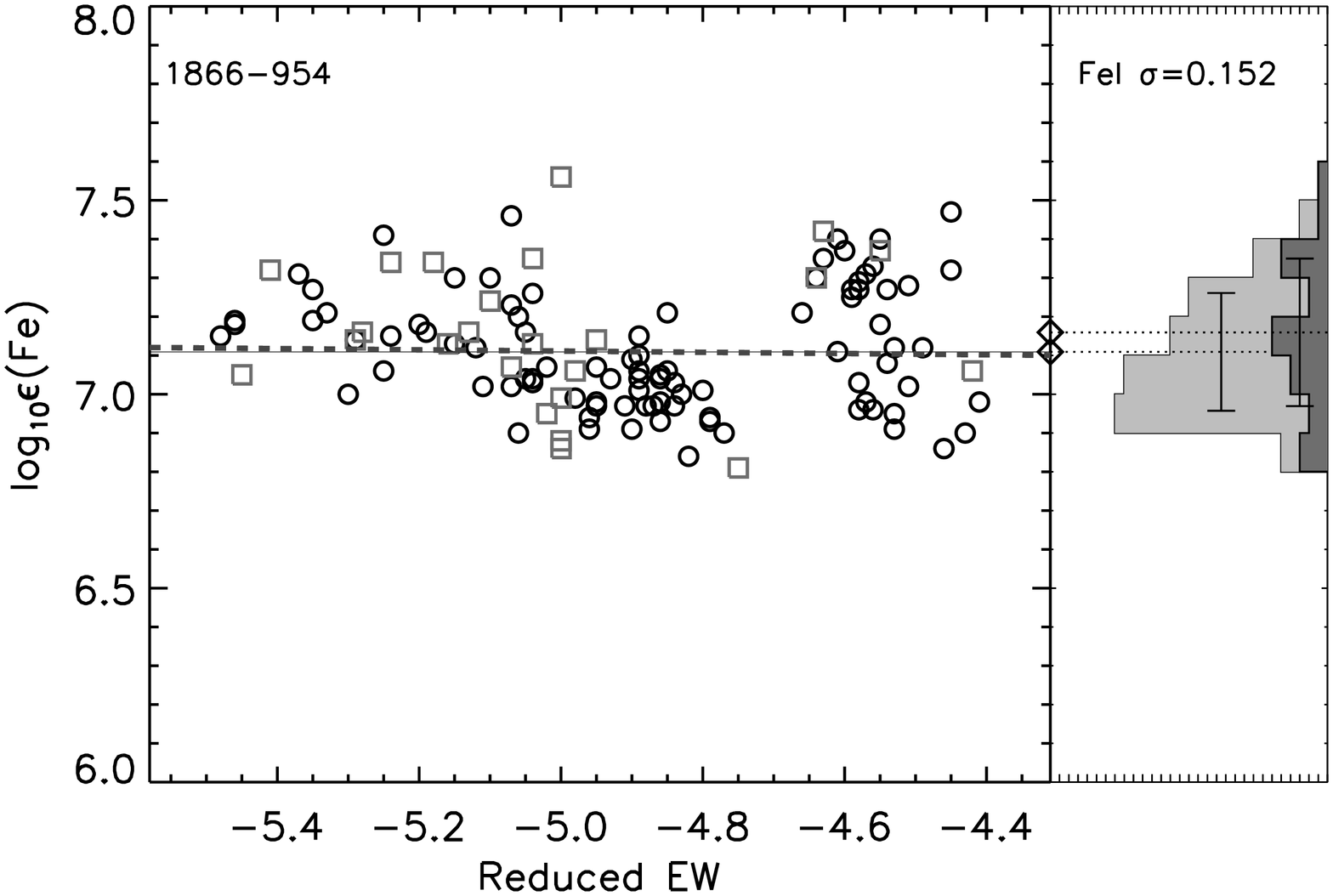}
\includegraphics[angle=90,scale=0.3]{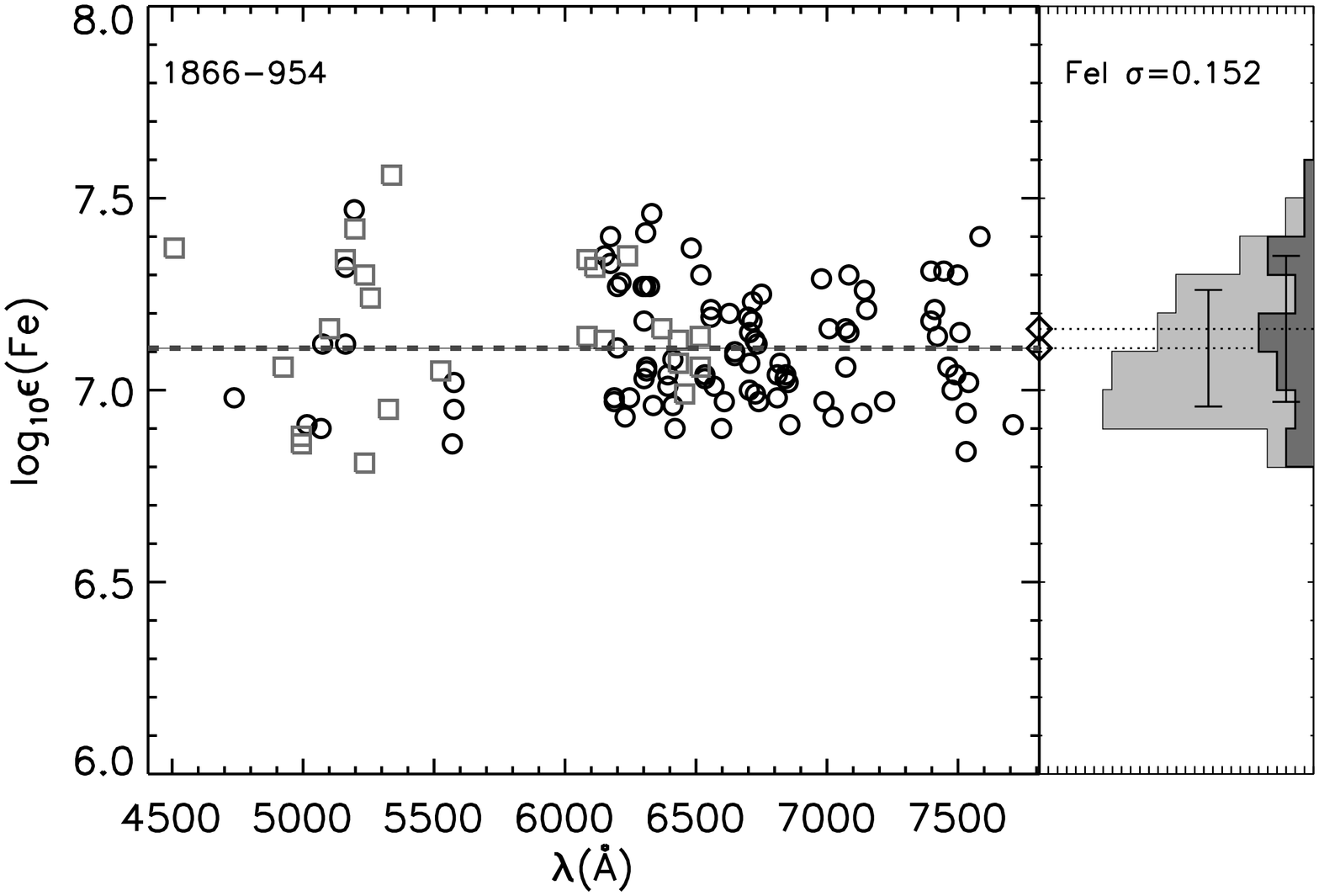}

\vskip 0.1cm
\caption{Examples of Fe abundance diagnostic plots, shown for star 954 in NGC 1866.  Fe I and Fe II lines are indicated by circles and squares, respectively. Solid lines show the mean abundance of the Fe I lines, which was used to constrain the stellar parameters.  Dashed lines show a linear least squares fit to the abundance of Fe I lines with EP, reduced EW, and wavelength, respectively.}
\label{fig: diagnostics} 
\end{figure}

\subsection{Cluster Stars}
\label{sec:stars}

To begin the analysis of the individual stars, we
determined initial atmospheric parameters for the cluster stars using the photometric data described in \textsection\ref{sec:data-stars}.  The reddening corrected absolute V magnitudes and $B-V$ colors that we used are listed in Table \ref{tab:lmcstars}.  With these colors we determined stellar temperatures using the empirical ($B-V$)-$T_{eff}$ calibration of \citet{alonso}.    Surface gravities are calculated according to the equation 
\begin{equation}\label{gravity} log~g = log~g_{\odot} + log~M/\msol - 
log~L/L_{\odot} + 4log~T_{eff}/T_{eff\odot} \end{equation}

assuming $T_{eff\odot}$=5777 K and log~$g_{\odot}$=4.44.  Bolometric
corrections are interpolated from the grids of
Kurucz,\footnote{Available from http://kurucz.harvard.edu/grids.html}
with M$_{\rm{bol}\odot}$=4.74.  We have assumed stellar masses of 1.8
$\msol$ for NGC 1978, 4.5 $\msol$ for NGC 1866, and 8.5 $\msol$ for
NGC 1711 and NGC 2100.  These masses were determined using the turnoff
masses of Teramo isochrones with appropriate ages and metallicities
for each cluster.  Stellar luminosities were calculated using the
distance moduli and $E(B-V)$ values listed in Table
\ref{tab:stellar_info}.  Initial microturbulent velocities ($\xi$)
were calculated as for our IL analysis \citep[see][]{mb08}, by
assuming a linear relationship between the $\xi$ of the Sun and
Arcturus.  As in the cluster IL analysis, we use the Kurucz ODFNEW
stellar atmospheres and the most recent (2010) version of MOOG
\citep{moog}.  We use the ODFNEW atmospheres instead of the AODFNEW
ones because we have determined that the stars are not significantly
enhanced in $\alpha$-elements, as presented in \citet{paper4}.

The initial values that we adopt for stellar mass, reddening, and
microturbulence are subject to the usual observational uncertainties;
we therefore constrain these parameters spectroscopically, as is
standard in abundance analysis of individual stars.  We first
iteratively adjust the effective temperature and microturbulence to
simultaneously obtain a solution with no dependence of Fe I abundance
on the excitation potential (EP) or reduced equivalent widths of
the lines.  On average, we find that the photometrically derived
T$_{eff}$ and $\xi$ values need to be adjusted by less than 100 K and
0.3 \kms to eliminate the dependence of abundance on EP and reduced
EW, respectively.  This is independent verification that the
microturbulence law that we employ in our IL analysis provides
reasonably accurate values for younger stars. However, in the case of
two very young supergiants, 2100-c12 and 2100-b22, our initial values
for $\xi$ were  significantly underestimated.  In these cases, we had to
adjust the $\xi$ by about 1.5 \kms, to $\xi$=3.3 \kms, in order to
eliminate the trend in Fe I abundance with reduced EW.  Because such
young supergiants are generally found to have $\xi\sim$3 \kms
\citep[e.g.][]{hill330}, we believe that these spectroscopically
determined values are reasonable.

In almost all cases, we find that the spectroscopically determined T$_{eff}$ and $\xi$ also result in solutions closer to ionization equilibrium, reducing the difference between abundances derived from Fe I and Fe II lines. This improvement is shown in Table \ref{tab:params}, where we tabulate Fe I and Fe II abundance results for our photometrically and spectroscopically determined parameters.  

 In two cases, 1978-730 and 1866-954,  we also adjust the surface gravity, log $g$,  to force ionization equilibrium for the Fe lines. In these two cases we find that adjusting the log $g$ not only results in a solution where neutral and ionized Fe  are closer to ionization equilibrium, but also neutral and ionized  Ti, Y, and Sc are closer to equilibrium.   In two other cases,  1866-1653 and 1711-988,   there are also   large differences  in abundance derived from Fe I and Fe II lines,  but we do not adjust the log $g$ in these cases because  we are unable to find a set of stellar parameters that simultaneously improve Fe, Ti, Y and Sc ionization equilibrium. 1866-1653 also has a particularly high line-to-line scatter  for Fe II of  $\sigma_{\rm{FeII}}$=0.40. In this case we keep the initial log $g$ value because of the uncertainty of the Fe II abundance.

\begin{figure*}
\centering
\includegraphics[scale=0.35]{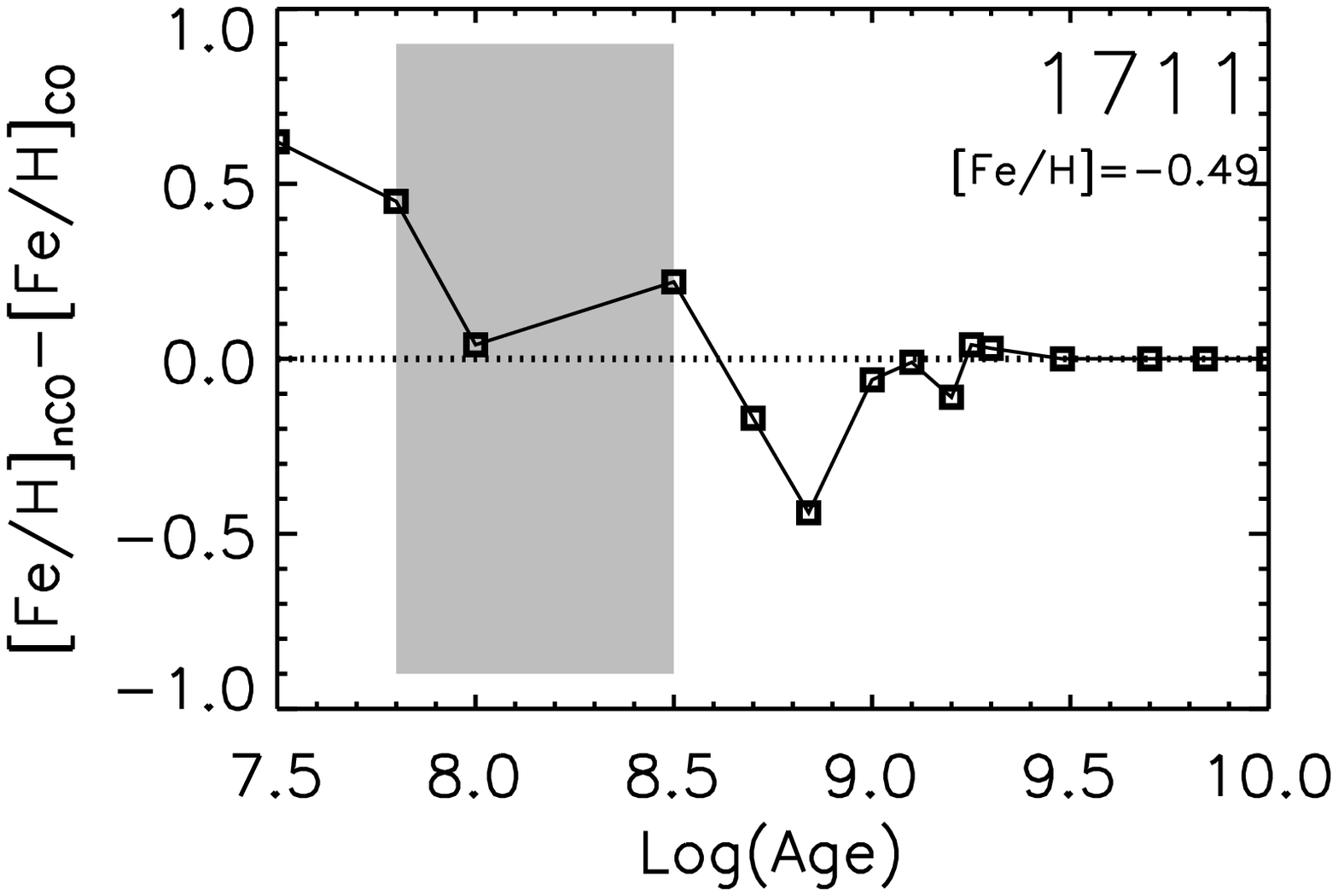}
\includegraphics[scale=0.35]{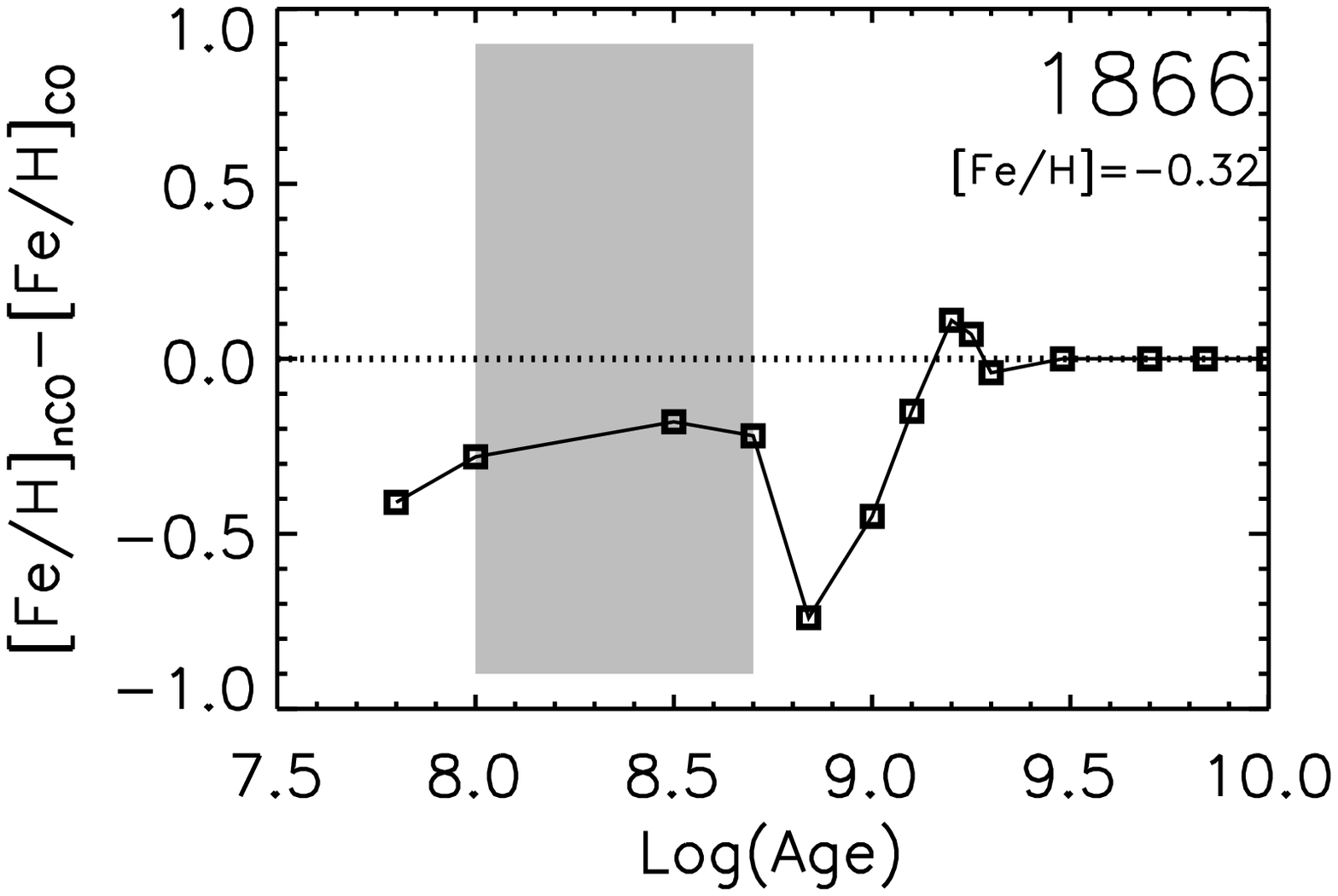}
\includegraphics[scale=0.35]{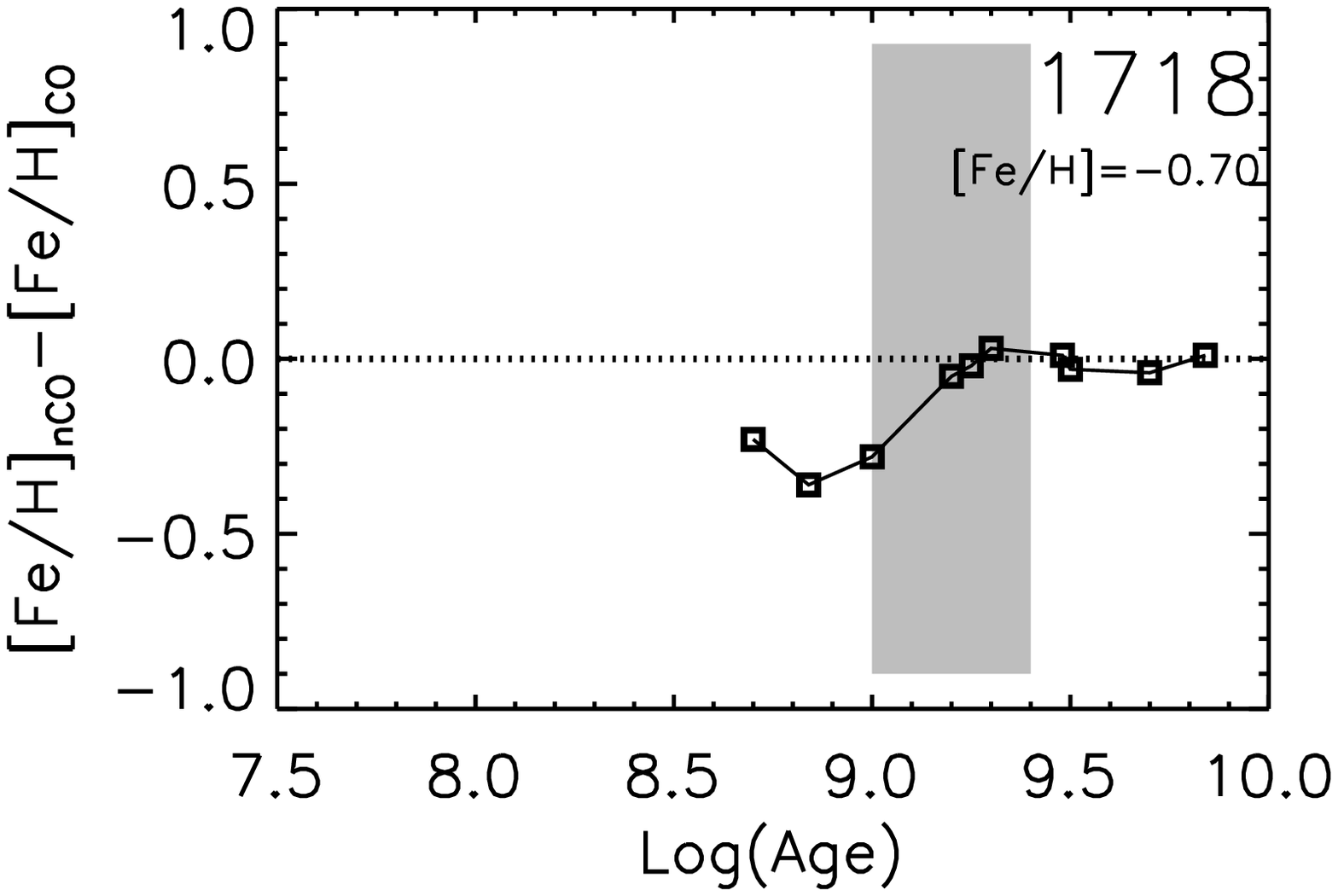}
\includegraphics[scale=0.35]{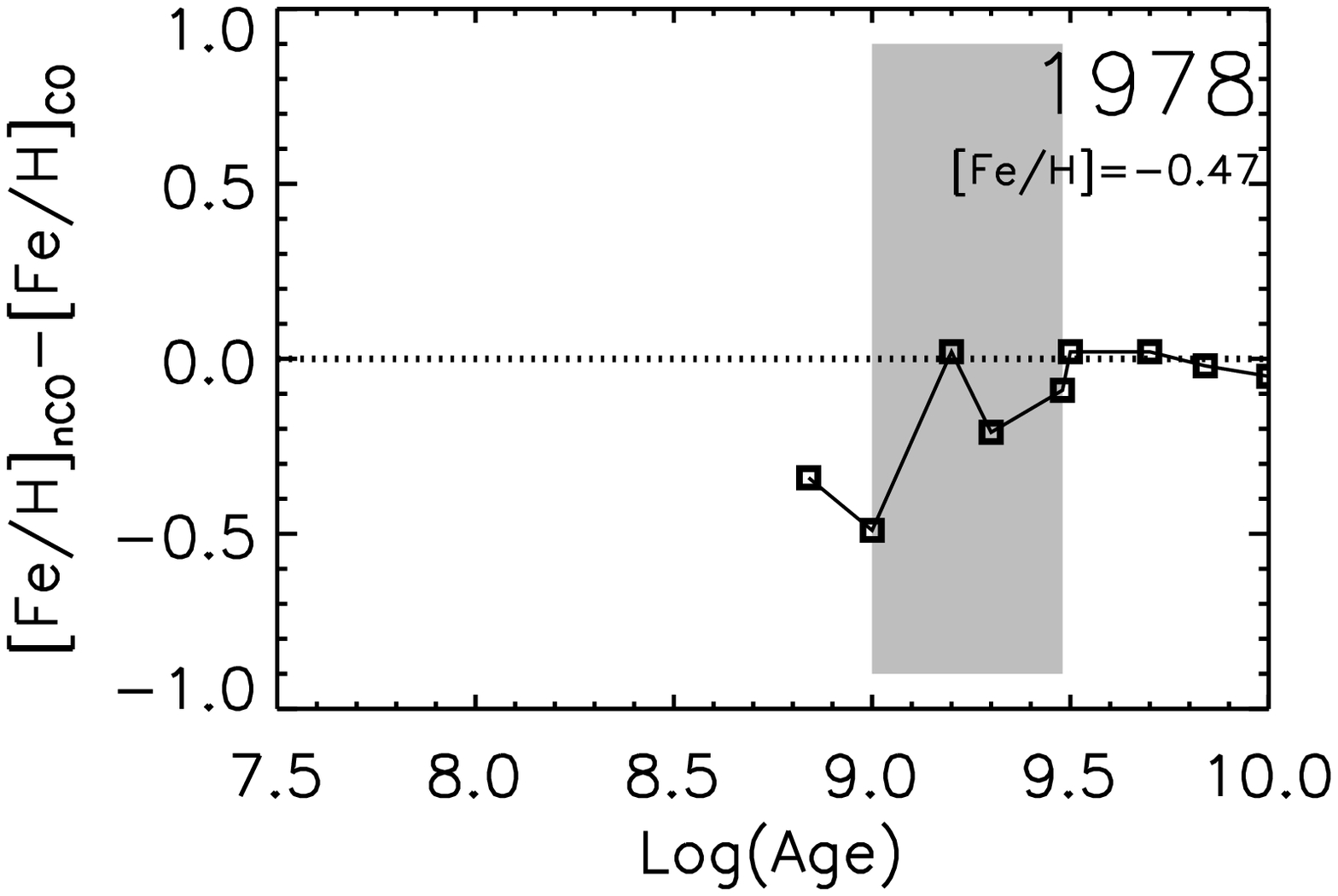}
\caption{Difference in derived  [Fe/H] for   isochrones with no C-OVER ([Fe/H]$_{\rm{nCO}}$)   (from this work) and with C-OVER ([Fe/H]$_{\rm{CO}}$) \citepalias[from ][]{paper3}  for IL analysis of  NGC 1711, NGC 1866, NGC 1718 and NGC 1978.  The age range with the best solutions in our IL analysis is highlighted.  Note that NGC 2100 is not included in this figure because the solutions between 0.06 and 1 Gyr do not converge due to this cluster's young age \citepalias[see ][for more details]{paper3}. }
\label{fig: feh} 
\end{figure*}

 Our final adopted stellar parameters and Fe abundance results are listed in Table \ref{tab:lmcstars}.  Our final model atmospheres are  adjusted so that the input [M/H] is identical to the derived Fe I abundance.  As an example of one of our solutions, we show the dependence of Fe abundance on  EP, reduced EW, and wavelength in Figure \ref{fig: diagnostics} for the star 954 in NGC 1866.  Little or no trend in the Fe abundance with wavelength indicates that our continuum placement between  orders is accurate and consistent.  
 
 We emphasize again that while the initial physical parameters used in
 the analysis of the spectra of stars (mass, effective
 temperature, and microturbulence) are drawn from the literature and
 those values can be affected by uncertainties such as reddening and
 photometric errors, the final values used for these parameters are
 spectroscopically constrained based on consistency in the iron
 abundance derived for lines over the full spectrum (i.e. with a wide
 range in wavelength, excitation potential, and reduced equivalent
 width).  The uncertainties in the initial values are therefore
 irrelevant as they are merely initial guesses.  What is relevant is
 the sensitivity of our solution to small errors in these
 iteratively-derived values.  We therefore determine the systematic
 uncertainties resulting from our choice of atmospheric parameters
 explicitly by adjusting one parameter at a time and noting the impact
 on the resulting abundance. The results are shown in Table
 \ref{tab:uncertainties}, where we list the uncertainty in the derived
 [Fe/H] for each star for an uncertainty in $T_{eff}$=$+150$ K,
 log~$g$=$-0.5$ dex, $\xi$=$+0.3$ \kms, and [M/H]=$+0.3$ dex. The
 total systematic error is typically between 0.1 and 0.2 dex.

To compare to our IL results, we average the [Fe/H] results for the individual stars.  For NGC 2100, we find that the star c2 has a derived [Fe/H] that is $\sim$ 0.4 dex more metal rich than the other two stars.  The three stars have a spread in radial velocity of 11 \kms, with c2 having the lowest velocity, so it is possible that c2 is not actually a cluster member. However the mean radial velocity for LMC field stars is most likely to be higher than the radial velocity for NGC 2100, not lower.  This star is also bluer and more luminous than the other two stars, so the stellar parameters may be more uncertain because it is in a different evolutionary stage. To be conservative, 2100-c2 is left out of the mean for the cluster.  The final mean [Fe/H] values are listed in Table \ref{tab:result}, along with previous abundance measurements by other authors from high resolution spectroscopy of individual stars.  There are no previous measurements of the Fe abundance of NGC 1711 from high resolution spectroscopy, so we list an estimate for the [Fe/H] measured from Str\"{o}mgren photometry by \cite{2000A&A...360..133D} for comparison.     We find that our [Fe/H] results agree within the uncertainties with the results found by other authors to $<$0.10 dex, with the exception of the [Fe/H] measured by \cite{2000A&A...364L..19H} for NGC 1978.     Our result for NGC 1978 does however agree with the  measurement of \cite{2008AJ....136..375M} who had a much larger sample of stars than \cite{2000A&A...364L..19H}.  We also note that \cite{1991AJ....101..515O} found a similar  [Fe/H] to ours for NGC 1978 using low resolution calcium triplet spectroscopy.  
We conclude that these comparisons demonstrate  that our stellar abundance analysis techniques are consistent with previous works.


\section{Results}
\label{sec:compare}

First we examine the general trends in the IL abundance results for the isochrones with and without C-OVER.  When we determine the best-fitting CMD in the IL analysis, we begin by identifying  one CMD for each age in our grid that has a self-consistent [Fe/H] solution.  We define self-consistent solutions as those where the derived [Fe/H] from the Fe I lines is the same as the [Fe/H] of the isochrone used to create the CMD.   We then identify the age range of the CMDs that produces the most stable [Fe/H] solution overall.  

 In comparing our results for isochrones with and without C-OVER, we can first look at how the self-consistent [Fe/H] solution changes as a function of the age of the CMD.  Figure \ref{fig: feh} shows  this [Fe/H] difference  for four clusters: NGC 1711, NGC 1866, NGC 1718, and NGC 1978.  Note that we can evaluate the behavior of the solutions with and without C-OVER for NGC 1718, but that without a sample of individual stars we cannot evaluate the agreement with our own fiducial Fe abundances.   In Figure \ref{fig: feh},  the region where the best age solutions are found is highlighted in each panel.   It is clear from this figure  that the clusters that are most affected by C-OVER are those younger than $\sim$1 Gyr.  As expected, we also find that the derived [Fe/H] for each cluster is unchanged when using the isochrones without C-OVER when isochrones older than $\sim$2.5 Gyr (log(Age)=9.40) are used, because there are no stars with masses $\gtrsim 1.1 ~\msol$ present (see \textsection \ref{sec:intro}).

\begin{figure*}
\centering
\includegraphics[scale=0.75]{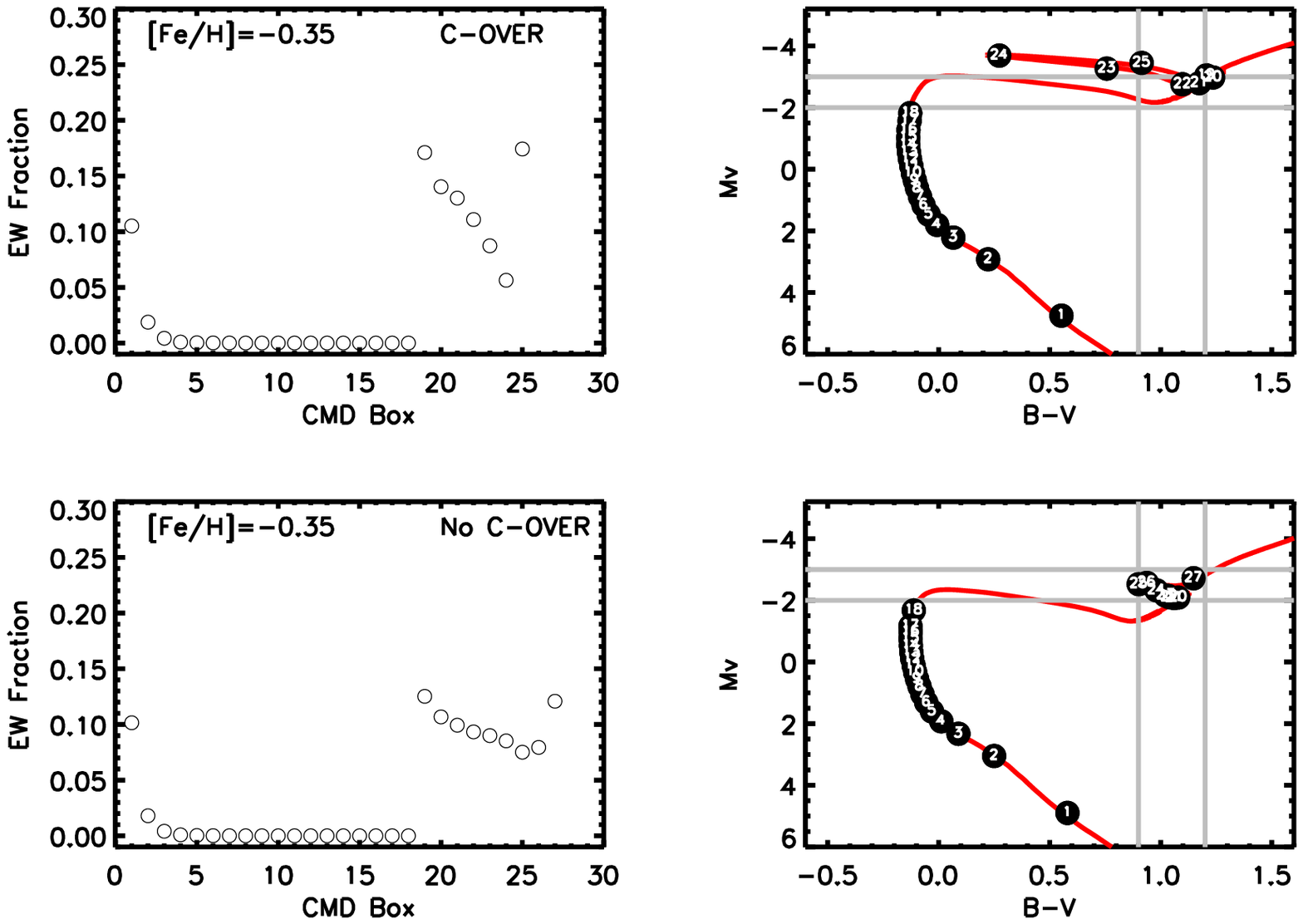}
\includegraphics[scale=0.75]{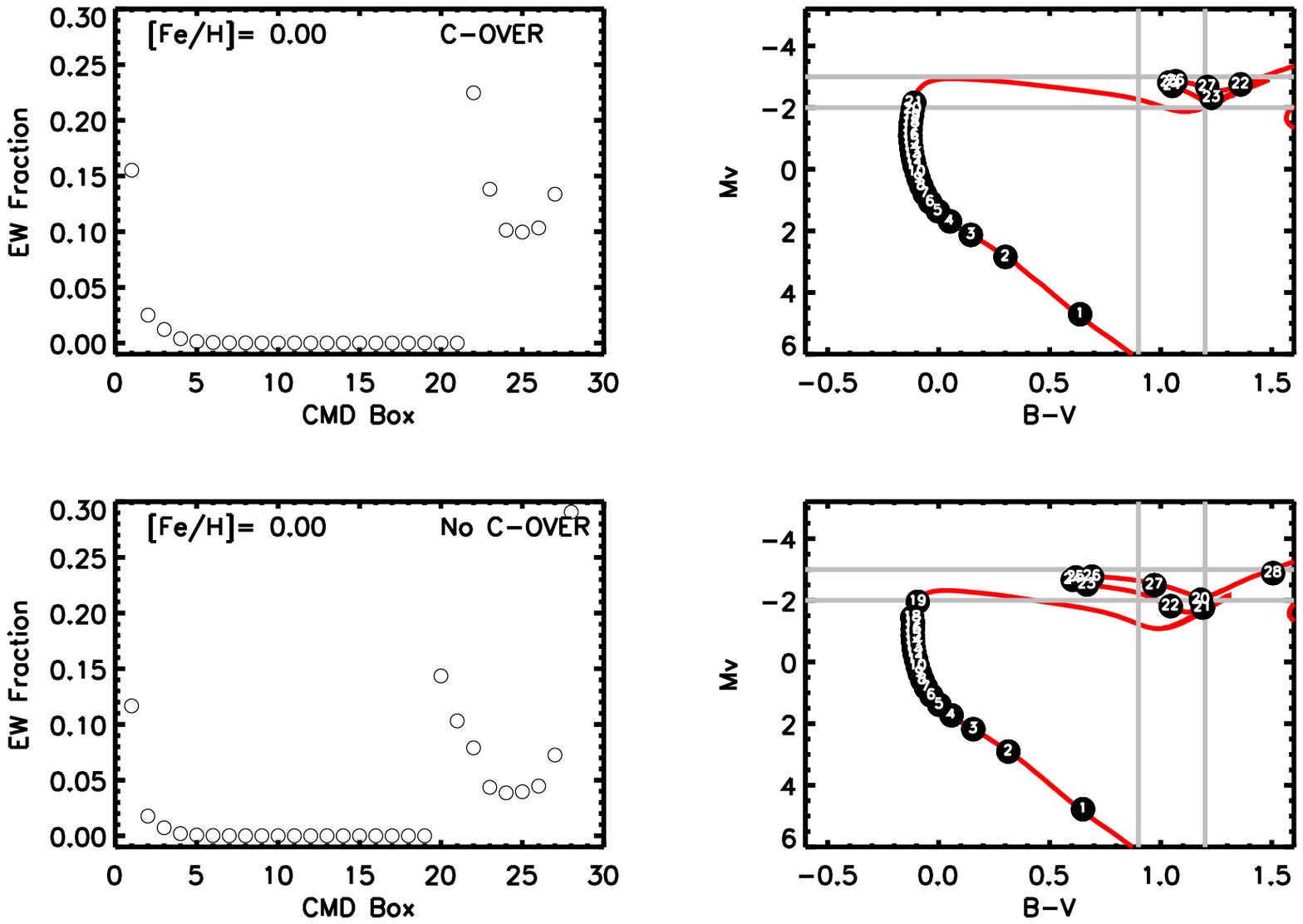}
\caption{ Left panels show the contribution of individual synthetic CMD boxes to the integrated light EW of the Fe I 6393 \rAA line (EP=2.43 eV), when different isochrones are used in constructing the CMDs. Right panels show the original isochrones in red, and the corresponding  CMD boxes from the left panels in black. Gray lines are shown to guide the eye to the position of the supergiant CMD boxes.  All isochrones have an age of 0.1 Gyrs.  Isochrones in the top four plots have [Fe/H]=$-0.35$, and isochrones in the bottom four plots have [Fe/H]=$0.00$.  }
\label{fig: fracew} 
\end{figure*}

 It is also interesting to note that the abundance is  sensitive to the amount of C-OVER because  90\% of the flux in a 100 Myr cluster, such as NGC 1866, is in the supergiant stars. This sensitivity was also illustrated in \citetalias{paper3} where we showed that the Fe lines are extremely sensitive to the color or temperature of the supergiant stars. To further illustrate the impact of these young stars,    in Figure \ref{fig: fracew}, we show the contribution of the individual synthetic CMD boxes to the IL EW for a single Fe line  for both [Fe/H]$=0$ and [Fe/H]$=-0.3$ for the two sets of isochrones.  Note that the best solution for NGC 1866 is [Fe/H]$=-0.3$ without C-OVER, and [Fe/H]=$0$ with C-OVER, resulting from the differences in the supergiant populations.  In Figure \ref{fig: fracew}, the left panels demonstrate that almost the entire EW comes from the supergiants, while  the right panels show that the CMDs differ most in the color of the supergiants.  In particular, the [Fe/H]$=-0.35$ isochrone with C-OVER has bluer supergiants than the isochrone with [Fe/H]=$-0.35$ and no C-OVER.  Note that only the latter provides a self-consistent [Fe/H] solution by our criterion discussed above.

Our final age and [Fe/H] solutions  for isochrones with and without C-OVER are listed in Table \ref{tab:abund_compare}.
We find that  the new IL results using isochrones without C-OVER show complex behavior when compared to the results obtained using isochrones with C-OVER.       In other words,  there is no constant offset that can be applied to the whole sample of clusters, because 
the new analysis results in higher [Fe/H] for NGC 1978, and NGC 1711, but lower [Fe/H] for NGC 1718 and NGC 1866.  As already mentioned, the differences tend to get larger as the age of the cluster decreases. This  means that the unpredictability in the magnitude and direction of the offset  is likely due to the fact that the properties of the giant stars are so susceptible to stochastic effects.

\begin{figure}
\centering
\includegraphics[trim = 0mm 100mm 0mm 0mm, clip,angle=90,scale=0.60]{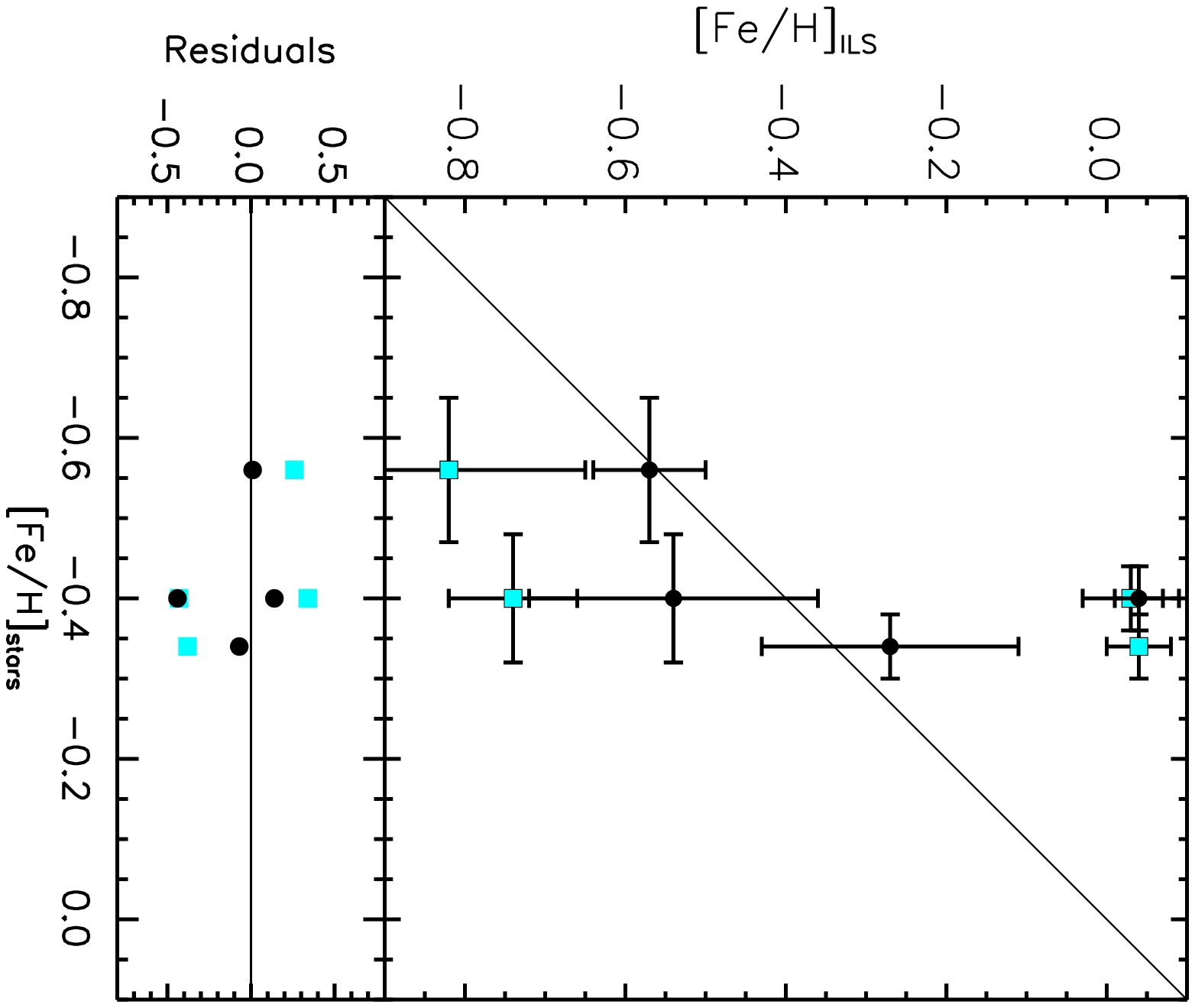}
\caption{IL spectra results for CMDs without C-OVER (black circles) and with C-OVER (cyan squares)  are  shown against the mean abundance obtained from individual stars. The bottom panel shows the residuals from the solid black 1:1 line. }
\label{fig:systematics}
\end{figure}

We can now use our stellar results to evaluate the accuracy of the isochrones as discussed in \textsection \ref{sec:intro}.
    Comparing the   results listed in  Tables \ref{tab:abund_compare} and \ref{tab:result}, we find that  the isochrones with no C-OVER, labeled ``canonical'' by the Teramo group, more closely match the results we obtain from the individual stars. This is also clear from 
    Figure \ref{fig:systematics}, where we plot the results using isochrones with and without C-OVER  against the abundance from individual  stars.    The bottom panel of Figure \ref{fig:systematics} shows the difference between the isochrone results and the individual star results.   The statistical scatter of the residuals for the isochrones without C-OVER, which are plotted as black circles, is 0.24 dex, while the scatter of the residuals for the isochrones with C-OVER, plotted as cyan squares, is much larger, at 0.40 dex.

        Figure \ref{fig:systematics} also shows that the largest difference in abundance determined from IL spectra is for the cluster with the highest [Fe/H], which is the youngest cluster, NGC 2100.  We note that our IL result for this cluster is the most uncertain, as it is at the youngest limit of what we can analyze using our current technique.      In \citetalias{paper3}, we  reported a lower limit for the [Fe/H] of NGC 2100, but it is obvious from this work that the cluster is more metal-poor than our limit, and perhaps an upper limit would have been more appropriate. We reported a lower limit because,  in general, if the age of the cluster is younger than the isochrone, we would underestimate the abundance because the stellar temperatures would be lower in our model than in reality.   However, in the case of very young clusters, age $<$50 Myr, the stochastic properties of the supergiants  can make the behavior unpredictable.    Therefore, we quote a  final [Fe/H] for NGC 2100 with higher uncertainties ($-0.4 <$ [Fe/H] $< +0.03$) rather than  a limit.  We note that while the [Fe/H] for NGC 2100 is uncertain,  the abundance ratios for all other elements are well constrained.  In \citet{paper4}, we discuss in more  detail the agreement found by  comparison of the abundance ratios obtained with IL analysis and individual stars.

 It is interesting  that \cite{2008AJ....136..375M} find that Teramo isochrones with some degree of C-OVER are needed to match the turnoff region of the CMD of  NGC 1978, and that we find  that isochrones without C-OVER produce a  more accurate [Fe/H] for this cluster in our IL analysis.  As we have emphasized already, it is not our intent in this work to use our IL spectroscopy results to calculate the appropriate magnitude of C-OVER that should be in stellar evolution models, or to comment on how the Teramo isochrones fit the observed CMDs of the clusters we have analyzed.   As inferred for the youngest clusters,  it is likely that differences in the most luminous RGB and AGB stars in the isochrones are the reason that there is a difference in the derived  [Fe/H], and that when averaged and  flux weighted  the isochrones without C-OVER more closely match the real stellar populations in this cluster.

 In conclusion, because clusters younger than 1 Gyr are most
 susceptible to stochastic stellar population effects, the abundance
 solutions for those clusters will be the most sensitive to including
 C-OVER in the CMDs.  This is because the supergiants, whose
 properties can change substantially when using C-OVER or not, have a
 strong influence on the solution, and therefore determine how much
 the C-OVER parametrization will affect the results.  Indeed, for
 clusters close to 0.05 Gyrs, the uncertainties in age and abundance
 determined by any method become much unavoidably impacted by both the
 stochastic effects of catching stars in the supergiant phases and the
 fundamental uncertainties in the simple stellar population (SSP) modeling.  We do not find a
 predictable offset in the solutions obtained without C-OVER; however
 we do find that smaller scatter is obtained in the comparison of
 abundances obtained from stars and integrated light when C-OVER is
 omitted from the models.  Therefore, for our IL abundance analysis
 purposes, the Teramo isochrones without C-OVER produce the most
 accurate [Fe/H] solutions.  More generally, in this and our companion
 papers, we find that measurements of age and abundance in young
 clusters should include uncertainties associated with the stochastic
 variations in supergiant population during any given observation, and
 also the uncertainties in the SSP models, of which C-OVER is one
 important consideration.

              
\section{Summary and Conclusions}
\label{sec:conclusions}

We have performed  high resolution Fe abundance analysis of individual stars in four LMC clusters in order to obtain fiducial Fe abundances in these clusters for comparison to Fe abundances that we measure using our high resolution integrated light spectra abundance analysis.  Our primary goal in this work is to determine the most appropriate convective overshooting parameter in the Teramo isochrone set for our  abundance analysis.  From these comparisons we conclude that Teramo isochrones without convective overshooting  result in Fe abundances that most closely match the results from individual stars for clusters with ages between 0.05 Gyr and 3 Gyr. In a separate paper, we present   abundances for over 20 additional elements that are measured in the cluster IL spectra, as well as in the spectra of the individual stars that are analyzed in this work.
  	
Finally, we note that	clusters in the LMC  have long been used to put constraints on stellar evolution and SSP models 	because the LMC is close enough that the stellar populations of star clusters are resolved, and the LMC contains rich clusters of a wide range in age ( 10's of Myrs to $\sim$12 Gyrs), unlike the Milky Way.    This age range is a critical regime for evaluating   evolutionary models, and   our IL abundance  analysis  method can provide a new and unique way to test stellar evolution models.  As this paper shows, IL analysis can be used to constrain how much C-OVER is allowable in the models.
 The IL analysis can also be used to compare the abundance results from stellar models computed by different groups.  For these tests to be most informative, it will be necessary to expand the sample of clusters with ages $<$3 Gyr that have both high resolution  IL and individual star abundance analyses available.

\acknowledgements
This research was supported by NSF grant AST-0507350. The authors thank the anonymous referee for helpful comments.  The authors thank A. McWilliam for careful reading of the manuscript and helpful suggestions. JEC thanks D. K. Lai for helpful discussions.



\clearpage

 \begin{deluxetable}{lllllcllll}
\scriptsize
\centering
\tablecolumns{10}
\tablewidth{0pc}
\tablecaption{Stellar Targets \label{tab:stellar_info}}
\tablehead{     \colhead{Name}  &\colhead{RA}  &\colhead{Dec} &
  \colhead{Run} & \colhead{T$_{\rm{exp}}$}  &    \colhead{S/N
    (pixel$^{-1}$)} & \colhead{$\lambda$} & \colhead{V} &\colhead{
    ($m-M$)$_{0}$} & \colhead{$E(B-V)$}  \\
 &\colhead{(J2000)} & \colhead{(J2000)}&  &\colhead{(s)} &
\colhead{(at 6500 \AA)} &\colhead{(\AA)} 
 }
\startdata
1978-737 & 82.17996 & $-$66.20839 & 2004 Oct & 14800 & 99 &  4800-9300 & 17.07  & 18.43 & 0.09\\ 

1978-730 & 82.21058 & $-$66.24611 &  2004 Oct  & 5800 & 81 &4800-9300
&16.85 &18.43 & 0.09 \\ 

\\
\hline
\\

1866-954 & 78.38208 &$-$65.46264 &2004 Oct  & 7530&     95 &  4100-9300 &15.97 &18.50 &0.06\\

1866-1653 & 78.43350 &$-$65.45370 &2003 Jan  & 7200 &     78 & 4500-7200 & 15.01 &18.50 &0.06\\

1866-1667 & 78.40430 &$-$65.45670 &2003 Jan  & 7200 &    54 & 4500-7200& 16.15 &18.50 &0.06\\

\\
\hline
\\

1711-831  &72.62896 & $-$69.97525 &2003 Nov & 7500&  118 &4700-8100&14.86 & 18.50&0.09\\

1711-988  &  72.66138 & $-$69.96819 &2003 Nov & 4000 &  97 &4700-8100&13.92 & 18.50&0.09\\

1711-1194 & 72.61679& $-$69.95275 & 2003 Nov & 7000 & 123&4700-8100 & 14.81 &18.50 &0.09\\

\\
\hline
\\

2100-b22 &  85.55092&   $-$69.20206             & 2003 Nov & 2100 &  115 &4700-8100&13.7  & 18.45 &0.24 \\
2100-c2  &  85.48508 &  $-$69.20747  & 2003 Nov & 2561 &  123 &4700-8100&13.0  & 18.45 &0.24 \\

2100-c12  & 85.53642 &$-$69.19544 & 2003 Nov&2100 &  120 &4700-8100& 13.8 &18.45 & 0.24\\
\enddata

\tablerefs{ Coordinates and magnitudes for each cluster were taken from the following catalogs: NGC 1978 \cite{1995A&AS..112..367W}, NGC 1866 \cite{1989ApJS...71...25B},  NGC 1711   \cite{1991A&AS...90..387S}, NGC 2100 \cite{1974A&AS...15..261R}. Adopted distance moduli and reddening values for the clusters were taken from the following:  NGC 1978 \cite{2006ApJ...645L..33F},  NGC 1866 \cite{mucc1866},   NGC 1711 \cite{2000A&A...360..133D}, NGC 2100 \cite{2000AJ....119.1748K}.}

\end{deluxetable}

 \clearpage

\LongTables
\scriptsize
\begin{deluxetable}{llllrrrrrrrrrrr}
\scriptsize
\tablecolumns{15}
\tablewidth{0pc}
\tablecaption{Line Parameters and Stellar Equivalent Widths \label{tab:linetable_stars}}
\tablehead{\colhead{Species} & \colhead{$\lambda$}   & \colhead{E.P.} & \colhead{log gf}   & \colhead{EW} & \colhead{EW} & \colhead{EW} & \colhead{EW} & \colhead{EW}& \colhead{EW}& \colhead{EW}& \colhead{EW}& \colhead{EW}&  \colhead{EW}&  \colhead{EW} \\ \colhead{} & \colhead{(\AA)} & \colhead{(eV)} &\colhead{}  & \colhead{(m\AA)}& \colhead{(m\AA)} & \colhead{(m\AA)} & \colhead{(m\AA)} & \colhead{(m\AA)} & \colhead{(m\AA)} & \colhead{(m\AA)} & \colhead{(m\AA)} & \colhead{(m\AA)} & \colhead{(m\AA)}  & \colhead{(m\AA)} \\
 \colhead{} & \colhead{} & \colhead{} &\colhead{} & \colhead{1978}&\colhead{1978}&\colhead{1866}&\colhead{1866}&\colhead{1866}&\colhead{1711} &\colhead{1711}&\colhead{1711}&\colhead{2100}&\colhead{2100} &\colhead{2100} \\
 \colhead{} & \colhead{} & \colhead{} &\colhead{} & \colhead{737}&\colhead{730}&\colhead{954}&\colhead{1653}&\colhead{1667}&\colhead{831} &\colhead{988}&\colhead{1194}&\colhead{c2}&\colhead{c12} &\colhead{b22}
}

\startdata

 Fe  I &4736.783 &  3.211 & -0.752 & \nodata & \nodata &185.1 & \nodata & \nodata & \nodata & 69.5 & \nodata &  \nodata&  \nodata &  \nodata \\
 Fe  I &5001.870 &  3.881 &  0.050 & \nodata & \nodata & \nodata & \nodata & \nodata &173.1 & \nodata & \nodata &  \nodata&  \nodata &  \nodata \\
 Fe  I &5001.870 &  3.881 &  0.050 & \nodata & \nodata & \nodata & \nodata & \nodata &153.5 & \nodata & \nodata &  \nodata&  \nodata &  \nodata \\
 Fe  I &5014.951 &  3.943 & -0.303 & \nodata & \nodata &148.2 & \nodata & \nodata & \nodata & \nodata & \nodata &  \nodata&  \nodata &  \nodata \\
 Fe  I &5068.771 &  2.940 & -1.041 & \nodata & \nodata &189.5 & \nodata & \nodata &195.6 & \nodata & \nodata &  \nodata&  \nodata &  \nodata \\
 Fe  I &5074.753 &  4.220 & -0.160 & \nodata & \nodata &150.6 & \nodata &132.0 &183.6 & \nodata & \nodata &  \nodata&  \nodata &  \nodata \\
 Fe  I &5162.281 &  4.178 &  0.020 & \nodata &141.8 &184.6 & \nodata & \nodata & \nodata & \nodata & \nodata &  \nodata&  \nodata &  \nodata \\
 Fe  I &5195.480 &  4.220 & -0.002 & \nodata & \nodata &185.4 & \nodata & \nodata &180.4 & \nodata & \nodata &  \nodata&  \nodata &  \nodata \\
 Fe  I &5367.476 &  4.415 &  0.443 & \nodata &132.6 &167.4 & \nodata &157.5 & \nodata & \nodata & \nodata &  \nodata&  \nodata &  \nodata \\
 Fe  I &5383.380 &  4.312 &  0.645 & \nodata & \nodata & \nodata & \nodata &159.4 & \nodata & \nodata & \nodata &  \nodata&  \nodata &  \nodata \\
 Fe  I &5389.486 &  4.415 & -0.410 & \nodata & \nodata & \nodata & \nodata & 99.0 & \nodata & 76.6 & \nodata &  \nodata&  \nodata &  \nodata \\
 Fe  I &5389.486 &  4.415 & -0.410 & \nodata & \nodata & \nodata & \nodata &140.8 & \nodata & \nodata & \nodata &  \nodata&  \nodata &  \nodata \\
 Fe  I &5569.631 &  3.417 & -0.500 & \nodata & \nodata &193.3 & \nodata & \nodata & \nodata & \nodata &205.4 &  \nodata&  \nodata &  \nodata \\
 Fe  I &5576.099 &  3.430 & -0.900 & \nodata & \nodata &171.0 & \nodata &158.5 &192.0 & \nodata &175.8 &  \nodata&  \nodata &  \nodata \\
Fe  I &5576.099 &  3.430 & -0.900 & \nodata & \nodata &165.2 & \nodata &147.4 &177.2 & \nodata &\nodata &  \nodata&  \nodata &  \nodata \\
 Fe  I &6151.623 &  2.180 & -3.330 &141.1 & \nodata &145.6 &173.9 & \nodata & \nodata & 26.1 & \nodata &  \nodata&  \nodata &  \nodata \\
 Fe  I &6173.341 &  2.220 & -2.863 & \nodata & \nodata &174.2 & \nodata &142.5 &190.8 & 42.8 & \nodata &  \nodata&  \nodata &  \nodata \\
Fe  I &6173.341 &  2.220 & -2.863 & \nodata & \nodata &168.8 & \nodata & \nodata & \nodata &  \nodata & \nodata &  \nodata&  \nodata &  \nodata \\
 Fe  I &6180.209 &  2.730 & -2.628 &142.8 &128.4    & \nodata &172.9 &126.3 &166.6 & \nodata & \nodata &  \nodata&  \nodata &  \nodata \\
Fe  I &6180.209 &  2.730 & -2.628 &138.6 &\nodata & \nodata & 136.5 &\nodata &\nodata & \nodata & \nodata &  \nodata&  \nodata &  \nodata \\
 Fe  I &6187.995 &  3.940 & -1.673 & 84.9 & \nodata & 83.8 &120.3 & 99.9 & 85.1 & 24.5 & \nodata &  \nodata& 136.0 & 158.3 \\
 Fe  I &6187.995 &  3.940 & -1.673 &102.6 & \nodata & 84.7 & 128.3 & 104.0 & \nodata & 27.8 & \nodata &  \nodata& \nodata& \nodata \\
 Fe  I &6200.321 &  2.610 & -2.386 &140.2 &   125.7     &163.0 & \nodata  &\nodata  &180.7 & 50.0 & \nodata &  \nodata&  \nodata & 240.3 \\
 Fe  I &6200.321 &  2.610 & -2.386 &\nodata&   \nodata     &153.5 & \nodata  &\nodata  &\nodata & 51.2 & \nodata &  \nodata&  \nodata & \nodata \\

 Fe  I &6213.437 &  2.220 & -2.490 & \nodata & \nodata &192.7 & \nodata & \nodata &195.8 & 70.4 & \nodata &  \nodata&  \nodata &  \nodata \\
 Fe  I &6219.287 &  2.200 & -2.428 & \nodata & \nodata & \nodata & \nodata &154.4 & \nodata & 81.2 & \nodata &  \nodata&  \nodata &  \nodata \\
 Fe  I &6229.232 &  2.830 & -2.821 &123.9 & \nodata &100.5 &147.1 & 99.8 &108.0 & \nodata& \nodata &  \nodata&  \nodata & 193.4 \\

Fe  I &6240.653 &  2.220 & -3.212 & \nodata & \nodata & \nodata & \nodata & \nodata &163.6 & 26.7 & \nodata &  \nodata&  \nodata &  \nodata \\
 Fe  I &6246.327 &  3.600 & -0.796 &137.9 &137.8 &167.6 &167.4 &149.7 &176.2 & 98.8 & \nodata &  \nodata&  \nodata &  \nodata \\
 Fe  I &6265.141 &  2.180 & -2.532 & \nodata & \nodata & \nodata & \nodata & \nodata & \nodata & 74.9 & \nodata &  \nodata&  \nodata &  \nodata \\
 Fe  I &6270.231 &  2.860 & -2.543 &130.9 &117.9 & \nodata & \nodata & \nodata &149.7 & \nodata & \nodata &  \nodata&  \nodata &  \nodata \\
 Fe  I &6297.799 &  2.220 & -2.669 & \nodata & \nodata &180.4 & \nodata & \nodata & \nodata & \nodata & \nodata &  \nodata&  \nodata &  \nodata \\
 Fe  I &6301.508 &  3.650 & -0.701 & \nodata &133.6 &176.9 & \nodata &154.9 & \nodata & \nodata &169.6 &  \nodata&  \nodata &  \nodata \\
 Fe  I &6301.508 &  3.650 & -0.701 & \nodata &149.9 & 166.5 & \nodata &\nodata & \nodata & \nodata &\nodata &  \nodata&  \nodata &  \nodata \\
 Fe  I &6307.854 &  3.640 & -3.270 & 35.7 & \nodata & 35.2 & \nodata & 37.4 & 27.2 & \nodata & 24.8 &  \nodata&  \nodata &  \nodata \\
Fe  I &6307.854 &  3.640 & -3.270 & 28.7 & \nodata &28.3 & \nodata &\nodata & 24.2 & \nodata &\nodata&  \nodata&  \nodata &  \nodata \\
 
Fe  I &6311.504 &  2.830 & -3.153 &129.0 & \nodata & 88.1 & \nodata &101.4 & \nodata & \nodata & \nodata &  \nodata&  \nodata & 192.9 \\
Fe  I &6311.504 &  2.830 & -3.153 &\nodata& \nodata & 88.6 & \nodata &\nodata & \nodata & \nodata & \nodata &  \nodata&  \nodata & \nodata \\

 Fe  I &6322.694 &  2.590 & -2.438 & \nodata & \nodata &163.6 & \nodata &154.1 &187.0 & 58.5 &167.0 &  \nodata&  \nodata &  \nodata \\
 Fe  I &6330.852 &  4.730 & -1.640 & 50.1 & 41.0 & 53.3 & \nodata & \nodata & 53.5 & \nodata & 52.4 &  \nodata&  75.0 &  73.0 \\
 Fe  I &6335.337 &  2.200 & -2.175 & \nodata & \nodata & \nodata & \nodata & \nodata & \nodata & 89.6 & \nodata &  \nodata&  \nodata &  \nodata \\
 Fe  I &6336.830 &  3.690 & -0.667 &147.8 &134.5 &168.2 & \nodata & \nodata &179.9 & 77.3 &176.9 &  \nodata&  \nodata &  \nodata \\
 Fe  I &6353.849 &  0.910 & -6.360 & 74.0 & 85.6 & \nodata &103.4 & \nodata & \nodata & \nodata & \nodata &  \nodata& 149.6 & 147.7 \\
 Fe  I &6355.035 &  2.840 & -2.328 & \nodata & \nodata & \nodata & \nodata & \nodata & \nodata & 40.5 & \nodata &  \nodata&  \nodata &  \nodata \\
 Fe  I &6380.750 &  4.190 & -1.366 & 97.5 & 80.7 & \nodata &128.0 & \nodata & \nodata & 34.8 & 90.3 &  \nodata&  \nodata & 154.5 \\
 Fe  I &6392.538 &  2.280 & -3.957 &106.1 &108.7 & 83.3 &125.3 & 62.3 & 95.5 & 11.4 & 95.5 &  \nodata&  \nodata & 170.4 \\
 Fe  I &6411.658 &  3.650 & -0.646 & \nodata & \nodata &175.0 & \nodata & \nodata &190.9 & 49.6 &198.7 &  \nodata&  \nodata &  \nodata \\
Fe  I &6411.658 &  3.650 & -0.646 & \nodata & \nodata &185.4 & \nodata & \nodata & 193.8 & 190.7  &\nodata&  \nodata&  \nodata &  \nodata \\
 Fe  I &6419.956 &  4.730 & -0.183 &125.0 & \nodata &108.6 &137.2 &101.0 & \nodata & 63.0 & 90.0 &  \nodata&  \nodata & 177.1 \\
 Fe  I &6475.632 &  2.560 & -2.929 &139.4 &139.5 & \nodata & \nodata &126.0 &168.7 & 60.2 &173.4 &  \nodata&  \nodata &  \nodata \\
 Fe  I &6481.878 &  2.280 & -2.985 & \nodata & \nodata &163.4 & \nodata & \nodata &199.0 & \nodata &196.6 &  \nodata&  \nodata &  \nodata \\
 Fe  I &6498.945 &  0.960 & -4.675 & \nodata & \nodata & \nodata & \nodata & \nodata & \nodata & 24.4 & \nodata &  \nodata&  \nodata &  \nodata \\
 Fe  I &6518.373 &  2.830 & -2.397 &144.6 & \nodata &149.0 & \nodata &123.4 &165.4 & \nodata &170.2 &  \nodata&  \nodata &  \nodata \\
 Fe  I &6533.940 &  4.540 & -1.360 & 53.8 & 57.2 & 59.4 & \nodata & \nodata & 64.7 & \nodata & 61.7 &113.7&  95.8 &  83.2 \\

Fe  I &6533.940 &  4.540 & -1.360 & 59.0 & 59.0 & 60.1 & \nodata & \nodata & \nodata  & \nodata & \nodata  & \nodata &  \nodata  &  \nodata  \\
 Fe  I &6556.806 &  4.790 & -1.720 & 28.4 & 62.4 & 29.2 & \nodata & 14.8 & 24.5 & \nodata & 26.3 &  \nodata&  \nodata &  \nodata \\
 Fe  I &6569.224 &  4.730 & -0.380 &134.1 & \nodata &103.4 &142.4 &126.2 & \nodata & \nodata & \nodata &  \nodata&  \nodata &  \nodata \\
 Fe  I &6571.180 &  4.290 & -2.950 & 27.5 & 38.8 & \nodata & \nodata & \nodata & \nodata & \nodata & \nodata &  \nodata&  \nodata &  \nodata \\
 Fe  I &6597.571 &  4.770 & -0.970 & 57.8 & 60.7 & 57.3 & 85.9 & 89.5 & 63.3 & 21.6 & \nodata &115.4& 120.9 &  \nodata \\
 Fe  I &6608.044 &  2.270 & -3.939 &133.1 & \nodata & 81.9 &148.2 & \nodata & \nodata & \nodata & \nodata &  \nodata&  \nodata &  \nodata \\
 Fe  I &6627.560 &  4.530 & -1.559 & 53.9 & 42.1 & 58.3 & 69.5 & 45.2 & \nodata & \nodata & \nodata & 84.5&  77.6 &  \nodata \\

 Fe  I &6646.966 &  2.600 & -3.917 & 71.6 & 58.6 & \nodata & 99.7 & 68.5 & \nodata & \nodata & \nodata &125.6& 111.6 &  \nodata \\
 Fe  I &6646.966 &  2.600 & -3.917 &\nodata & \nodata & \nodata & \nodata &\nodata & \nodata & \nodata & \nodata &\nodata & 118.8&  \nodata \\

 Fe  I &6648.121 &  1.010 & -5.730 &106.7 & 98.1 & 85.2 &141.6 & 58.5 & \nodata & \nodata & \nodata &\nodata& \nodata &  \nodata \\
 Fe  I &6648.121 &  1.010 & -5.730 & \nodata & \nodata  & 84.7 &\nodata & \nodata & \nodata & \nodata & \nodata &\nodata& \nodata &  \nodata \\

 Fe  I &6653.911 &  4.140 & -2.447 & 39.8 & 34.0 & \nodata & 53.0 & \nodata & \nodata & \nodata & \nodata &  \nodata&  60.8 &  \nodata \\
 Fe  I &6699.136 &  4.590 & -2.117 & 16.3 & \nodata & 23.0 & \nodata & \nodata & 23.0 & \nodata & 20.7 &  \nodata&  \nodata &  \nodata \\
 Fe  I &6703.576 &  2.760 & -3.059 &102.3 &124.4 & 98.7 & \nodata & \nodata & \nodata & 16.2 & \nodata &  \nodata&  \nodata &  \nodata \\
 Fe  I &6704.500 &  4.200 & -2.587 & 22.0 & \nodata & 22.1 & \nodata & 12.0 & 27.0 & \nodata & 26.7 &  \nodata&  \nodata &  \nodata \\
 Fe  I &6705.105 &  4.610 & -1.060 & 59.2 & \nodata & 75.2 & \nodata &105.1 & 73.3 & \nodata & \nodata &  \nodata&  96.6 &  \nodata \\
 Fe  I &6710.323 &  1.480 & -4.807 &116.3 &139.0 & \nodata & \nodata &119.2 &158.2 & \nodata &159.6 &  \nodata&  \nodata &  \nodata \\
 Fe  I &6713.745 &  4.790 & -1.479 & \nodata & \nodata & 42.2 & \nodata & 34.7 & 40.7 & \nodata & \nodata &  \nodata&  54.3 &  \nodata \\
 Fe  I &6715.386 &  4.590 & -1.540 & 65.5 & \nodata & 57.4 & \nodata & 49.5 & \nodata & \nodata & \nodata &  \nodata& 117.2 &  \nodata \\
 Fe  I &6725.364 &  4.100 & -2.227 & 49.7 & \nodata & 47.1 & \nodata & 44.5 & 51.7 & \nodata & \nodata &  \nodata&  91.2 &  \nodata \\
 Fe  I &6726.673 &  4.590 & -1.087 & 63.5 & \nodata & 71.0 & \nodata & \nodata & 67.7 & \nodata & \nodata &  \nodata&  \nodata &  \nodata \\
 Fe  I &6733.153 &  4.620 & -1.479 & 46.0 & \nodata & 51.5 & \nodata & \nodata & 50.6 & \nodata & 50.2 &  \nodata&  68.5 &  \nodata \\
 Fe  I &6739.524 &  1.560 & -4.801 &110.3 & \nodata & 89.8 & \nodata & 63.3 &103.1 & \nodata &109.6 &  \nodata&  \nodata &  \nodata \\
 Fe  I &6750.164 &  2.420 & -2.592 & \nodata & \nodata &173.1 & \nodata &159.0 & \nodata & 50.8 &200.3 &  \nodata&  \nodata &  \nodata \\
 Fe  I &6806.856 &  2.730 & -2.633 &113.0 &115.2 & \nodata &140.8 &112.5 &111.9 & \nodata &111.4 &195.6&  \nodata &  \nodata \\

 Fe  I &6806.856 &  2.730 & -2.633 & \nodata & \nodata & \nodata & \nodata&125.4 & \nodata& \nodata & \nodata &185.4&  \nodata &  \nodata \\

 Fe  I &6810.267 &  4.590 & -0.992 & 84.8         &106.1     & 80.3   & 92.5    & \nodata  & 81.2 & 26.3 & \nodata &122.2& 133.8 &  \nodata \\
 Fe  I &6810.267 &  4.590 & -0.992 &  \nodata   & \nodata &76.2 & \nodata  & \nodata  & 75.0 &  \nodata & \nodata & \nodata &  \nodata  &  \nodata \\


 Fe  I &6820.374 &  4.620 & -1.214 & 70.3 & 69.2 & 65.5 & \nodata & \nodata & 68.8 & \nodata & 72.1 &122.5& 125.3 &  \nodata \\
 Fe I &6828.596 &  4.640 & -0.843 & \nodata & \nodata & \nodata & \nodata & \nodata & \nodata & \nodata & \nodata &152.7&  \nodata &  \nodata \\
 Fe  I &6839.835 &  2.560 & -3.378 &106.6 &114.4 & 98.7 &126.7 & 88.1 & \nodata & 15.8 & \nodata &  \nodata&  \nodata &  \nodata \\
 Fe  I &6841.341 &  4.610 & -0.733 & \nodata & \nodata & \nodata &138.4 & 81.7 & \nodata & \nodata & \nodata &  \nodata&  \nodata &  \nodata \\
 Fe  I &6842.689 &  4.640 & -1.224 & 62.1 & 49.2 & 61.4 & 81.3 & 45.7 & 71.2 & 27.2 & \nodata &  \nodata&  \nodata &  \nodata \\
 Fe  I &6843.655 &  3.650 & -0.863 & \nodata & \nodata & \nodata & \nodata & 59.3 & \nodata & \nodata & \nodata &  \nodata&  \nodata &  \nodata \\
 Fe  I &6851.652 &  1.600 & -5.247 & 73.6 & 85.0 & 57.9 & \nodata & 54.3 & \nodata & \nodata & \nodata &169.0&  \nodata &  \nodata \\
 Fe  I &6855.723 &  4.390 & -1.747 & 56.5 & 65.8 & \nodata & \nodata & 40.2 & \nodata & \nodata & \nodata &  \nodata&  \nodata &  \nodata \\
 Fe  I &6858.155 &  4.590 & -0.939 & 72.2 & 79.6 & 75.1 & \nodata & 69.3 & \nodata & 31.9 & \nodata &  \nodata&  \nodata &  \nodata \\
 Fe  I &6916.686 &  4.150 & -1.359 & \nodata & \nodata & \nodata & \nodata & \nodata & \nodata & \nodata &134.8 &  \nodata&  \nodata & 177.2 \\
 Fe  I &6960.330 &  4.570 & -1.907 & \nodata & \nodata & \nodata & 52.2 & \nodata & \nodata & \nodata & \nodata &  \nodata& 101.7 &  85.4 \\
 Fe  I &6978.862 &  2.480 & -2.465 & \nodata & \nodata &182.0 & \nodata & \nodata & \nodata & \nodata & \nodata &  \nodata&  \nodata &  \nodata \\
 Fe  I &6988.533 &  2.400 & -3.519 &119.8 &118.7 &100.3 & \nodata & 83.3 &123.2 & \nodata &132.1 &  \nodata&  \nodata &  \nodata \\
 Fe  I &7007.976 &  4.180 & -1.929 & 66.8 & 67.1 & 62.4 & \nodata & 38.5 & 67.9 & \nodata & \nodata &  \nodata&  \nodata &  \nodata \\
 Fe  I &7022.957 &  4.190 & -1.148 & 99.9 & 91.1 & 97.0 & \nodata & \nodata & 94.8 & 40.3 & 97.7 &  \nodata&  \nodata & 157.2 \\
 Fe  I &7024.644 &  4.540 & -1.106 &107.3 &102.6 & \nodata & \nodata & \nodata & \nodata & \nodata & \nodata &  \nodata&  \nodata &  \nodata \\
 Fe  I &7038.220 &  4.220 & -1.214 &100.9 & 89.9 & \nodata & \nodata & \nodata & \nodata & 36.9 & \nodata &  \nodata&  \nodata &  \nodata \\
 Fe  I &7068.423 &  4.070 & -1.319 &141.8 & \nodata & \nodata & \nodata &117.2 & \nodata & 28.7 & \nodata &  \nodata&  \nodata &  \nodata \\
 Fe  I &7071.866 &  4.610 & -1.627 & 73.8 & \nodata & 39.9 & 62.5 & 60.0 & 45.0 & \nodata & 48.2 & 97.5&  87.2 & 117.9 \\
 Fe  I &7072.800 &  4.070 & -2.767 & \nodata & \nodata & \nodata & \nodata & \nodata & \nodata & \nodata & \nodata &  \nodata&  93.3 &  \nodata \\
 Fe  I &7083.394 &  4.910 & -1.327 & 37.1 & \nodata & 50.2 & 59.7 & 44.3 & 42.6 & \nodata & 36.8 &  \nodata&  \nodata &  \nodata \\
 Fe  I &7090.390 &  4.230 & -1.109 & \nodata & \nodata & \nodata & \nodata &118.2 & \nodata & 36.3 & \nodata &  \nodata&  \nodata &  \nodata \\
 Fe  I &7132.985 &  4.060 & -1.635 & \nodata & \nodata & 77.7 &132.9 & 77.1 & 79.9 & \nodata & 79.2 &  \nodata&  \nodata &  \nodata \\
 Fe  I &7142.517 &  4.930 & -1.017 & \nodata & \nodata & 64.9 &103.7 & 73.9 & \nodata & \nodata & \nodata &  \nodata&  \nodata &  \nodata \\
 Fe  I &7151.464 &  2.480 & -3.657 &124.8 & \nodata &100.9 & \nodata & \nodata & 35.3 & \nodata & \nodata &  \nodata&  \nodata &  \nodata \\
 Fe  I &7219.680 &  4.060 & -1.617 & 77.5 & 99.0 & 80.5 & \nodata & 75.9 & 85.2 & 18.1 & \nodata &136.3& 127.8 &  \nodata \\
 Fe  I &7396.526 &  4.990 & -1.567 & \nodata & \nodata & 31.7 & \nodata & \nodata & 23.5 & \nodata & \nodata &  \nodata&  \nodata &  \nodata \\
 Fe  I &7396.526 &  4.990 & -1.567 & \nodata & \nodata & 25.9 & \nodata & \nodata & \nodata& \nodata & \nodata &  \nodata&  \nodata &  \nodata \\

 Fe  I &7411.162 &  4.280 & -0.287 &137.7 &138.1 &161.8 & \nodata & \nodata &197.4 & 85.0 & \nodata &149.8& 135.4 &  \nodata \\
 Fe  I &7421.560 &  4.640 & -1.727 & 37.0 & 33.6 & 37.8 & \nodata & \nodata & 34.5 & \nodata & \nodata &  \nodata&  47.3 &  41.2 \\
 Fe  I &7445.758 &  4.260 &  0.053 & \nodata & \nodata &199.2 & \nodata & \nodata & \nodata & \nodata & \nodata &  \nodata&  \nodata &  \nodata \\
 Fe  I &7454.004 &  4.190 & -2.337 & 48.6 & 33.3 & \nodata & \nodata & \nodata & \nodata & \nodata & \nodata & 80.3&  70.9 &  \nodata \\
 Fe  I &7461.527 &  2.560 & -3.507 &116.5 &112.0 & 95.9 & \nodata & \nodata & \nodata & \nodata & \nodata &  \nodata&  \nodata &  \nodata \\
 Fe  I &7477.595 &  3.880 & -2.560 & 42.8 & 38.7 & 37.8 & \nodata & \nodata & \nodata & \nodata & \nodata &  \nodata&  68.8 &  56.8 \\
 Fe  I &7491.652 &  4.280 & -1.067 & \nodata & \nodata &103.4 & \nodata & \nodata & \nodata & 39.9 & \nodata &  \nodata&  \nodata &  \nodata \\
 Fe  I &7498.535 &  4.140 & -2.177 & 68.2 & 65.4 & 60.1 & \nodata & \nodata & \nodata & \nodata & \nodata &  \nodata&  87.4 &  85.8 \\
 Fe  I &7506.030 &  5.060 & -1.230 & 48.3 & 40.8 & \nodata & \nodata & \nodata & \nodata & \nodata & \nodata &107.5&  98.5 &  \nodata \\
 Fe  I &7507.273 &  4.410 & -1.107 & 94.3 & 83.7 & 97.5 & \nodata & \nodata & \nodata & 35.7 & \nodata &  \nodata&  \nodata & 151.9 \\
 Fe  I &7531.153 &  4.370 & -0.557 &117.9 &139.1 &120.9 & \nodata & \nodata &134.6 & 62.5 & \nodata &  \nodata&  \nodata &  \nodata \\
 Fe  I &7540.444 &  2.730 & -3.777 & 61.0 & 76.7 & 58.1 & \nodata & \nodata & 70.0 & \nodata & \nodata &165.5& 116.3 & 125.6 \\
 Fe  I &7583.790 &  3.018 & -1.885 & \nodata & \nodata &186.4 & \nodata & \nodata &192.8 & \nodata & \nodata &  \nodata&  \nodata &  \nodata \\
 Fe  I &7710.363 &  4.220 & -1.113 & \nodata & \nodata & 98.1 & \nodata & \nodata & \nodata & \nodata & \nodata &  \nodata&  \nodata & 155.9 \\
 Fe  I &7941.085 &  3.274 & -2.286 & \nodata & 97.9 &166.4 & \nodata & \nodata & \nodata & \nodata & \nodata &  \nodata&  \nodata &  \nodata \\

&\\ 

Fe II &4508.288 &  2.856 & -2.440 & \nodata & \nodata &127.8 & \nodata & \nodata & \nodata & \nodata & \nodata &  \nodata&  \nodata &  \nodata \\
 Fe II &4831.126 &  0.000 & -4.890 & \nodata & \nodata & \nodata & \nodata &158.6 & \nodata & \nodata & \nodata &  \nodata&  \nodata &  \nodata \\
 Fe II &4833.197 &  2.657 & -4.640 & \nodata & \nodata & \nodata & \nodata & 33.9 & \nodata & \nodata & \nodata &  \nodata&  \nodata &  \nodata \\
 Fe II &4833.865 &  2.844 & -5.110 & \nodata & \nodata & \nodata & \nodata & 36.9 & \nodata & \nodata & \nodata &  \nodata&  \nodata &  \nodata \\
 Fe II &4923.927 &  2.891 & -1.260 &144.3 &148.9 &188.1 & \nodata & \nodata & \nodata & \nodata & \nodata &  \nodata&  \nodata &  \nodata \\
 Fe II &4993.358 &  2.807 & -3.620 & 88.8 & \nodata & 49.7 & \nodata & \nodata & 56.3 & \nodata & \nodata &  \nodata&  \nodata &  \nodata \\
 Fe II &5036.920 &  3.017 & -4.670 & \nodata & \nodata & \nodata & \nodata & \nodata & \nodata & 41.8 & \nodata &  \nodata&  \nodata &  \nodata \\
 Fe II &5100.664 &  2.807 & -4.170 & \nodata & \nodata & 37.6 & \nodata & \nodata & 37.3 & \nodata & \nodata &  \nodata&  \nodata &  \nodata \\
 Fe II &5120.352 &  2.828 & -4.240 & \nodata & \nodata & \nodata & \nodata & \nodata & \nodata & 81.5 & \nodata &  \nodata&  \nodata &  \nodata \\
 Fe II &5132.669 &  2.807 & -4.080 & \nodata & \nodata & \nodata & \nodata & \nodata & \nodata & 92.5 & \nodata &  \nodata&  \nodata &  \nodata \\
 Fe II &5146.127 &  2.828 & -3.910 & \nodata & \nodata & \nodata & \nodata & \nodata & \nodata & 98.0 & \nodata &  \nodata&  \nodata &  \nodata \\

Fe II &5154.409 &  2.844 & -4.130 & 57.9 & \nodata & 50.3 & \nodata & \nodata & \nodata & \nodata & \nodata &  \nodata&  \nodata &  \nodata \\
 Fe II &5161.184 &  2.856 & -4.470 & \nodata & \nodata & 30.0 & \nodata & \nodata & \nodata & 44.4 & \nodata &  \nodata&  \nodata &  \nodata \\
 Fe II &5197.577 &  3.230 & -2.220 & \nodata & \nodata &121.9 & \nodata & \nodata & \nodata & 44.5 & \nodata &  \nodata&  \nodata &  \nodata \\
 Fe II &5234.625 &  3.221 & -2.180 & 88.9 & 93.6 & 93.7 &111.2 &127.5 & \nodata & \nodata & \nodata &  \nodata&  \nodata & 171.6 \\
 Fe II &5234.625 &  3.221 & -2.180 & 96.3 & \nodata &  118.6 & 109.6 &\nodata& \nodata & \nodata & \nodata &  \nodata&  \nodata & \nodata \\
 Fe II &5256.938 &  2.891 & -4.060 & \nodata & \nodata & 41.8 & \nodata & \nodata & \nodata & 94.6 & \nodata &  \nodata&  \nodata &  \nodata \\
 Fe II &5264.812 &  3.230 & -3.130 & \nodata & \nodata & \nodata & 49.6 & \nodata & \nodata & \nodata & \nodata &  \nodata&  \nodata &  \nodata \\
 Fe II &5325.553 &  3.221 & -3.160 & 44.4 & \nodata & 51.1 & 35.8 &109.6 & \nodata & \nodata & \nodata &  \nodata&  84.4 &  \nodata \\
 Fe II &5337.732 &  3.230 & -3.720 & \nodata & 29.7 & 53.3 & 49.1 & 36.7 & \nodata & \nodata & \nodata &  \nodata&  66.2 &  58.8 \\
 Fe II &5414.073 &  3.221 & -3.580 & 44.8 & \nodata & \nodata & \nodata & 62.2 & \nodata & \nodata & \nodata &  \nodata& 113.2 &  \nodata \\
 Fe II &5425.257 &  3.199 & -3.220 & 40.9 & 34.3 & \nodata & 42.1 & 50.8 & 42.4 & \nodata & \nodata & 99.5&  55.8 &  55.5 \\
 Fe II &5525.125 &  3.267 & -3.970 & \nodata & 30.1 & 19.6 & 28.0 & \nodata & \nodata & 52.8 & \nodata &  \nodata&  57.7 &  44.9 \\
 Fe II &5591.368 &  3.267 & -4.440 & \nodata & \nodata & \nodata & \nodata & \nodata & \nodata & 29.6 & 10.3 &  \nodata&  30.2 &  \nodata \\
 Fe II &5591.368 &  3.267 & -4.440 & \nodata & \nodata & \nodata & \nodata & \nodata & \nodata & 27.4 & 17.9 &  \nodata&  24.2 &  \nodata \\
 Fe II &5627.497 &  3.387 & -4.100 & \nodata & \nodata & \nodata & \nodata & \nodata & \nodata & 46.4 & \nodata  &  \nodata&  \nodata &  \nodata \\
 Fe II &5813.677 &  5.571 & -2.510 & \nodata & \nodata & \nodata & \nodata & \nodata & \nodata & 18.4 & \nodata &  \nodata&  \nodata &  \nodata \\
 Fe II &5991.376 &  3.153 & -3.540 & 35.3 & 37.0 & \nodata & \nodata & \nodata & \nodata & \nodata & 37.2 &  \nodata&  63.4 &  \nodata \\
 Fe II &6084.111 &  3.199 & -3.790 & 20.9 & 23.7 & 31.0 & \nodata & 55.2 & 29.5 & 82.2 & \nodata & 91.2&  \nodata &  41.0 \\
 Fe II &6113.322 &  3.221 & -4.140 & \nodata & \nodata & 24.0 & \nodata & 54.3 & \nodata & 56.2 & \nodata &  \nodata&  \nodata &  \nodata \\
 Fe II &6147.741 &  3.889 & -2.690 & \nodata & \nodata & \nodata & \nodata &103.2 & \nodata & \nodata & \nodata &  \nodata&  \nodata &  \nodata \\
 Fe II &6149.258 &  3.889 & -2.690 & 50.3 & \nodata & 42.2 & 54.5 & \nodata & \nodata & 87.0 & \nodata &  \nodata&  \nodata &  \nodata \\
 Fe II &6238.392 &  3.889 & -2.600 & 50.7 & \nodata & 56.7 & 79.3 & \nodata & \nodata & \nodata & \nodata &  \nodata&  \nodata &  \nodata \\
 Fe II &6369.462 &  2.891 & -4.110 & 35.1 & 30.0 & 33.8 & \nodata & \nodata & \nodata & 82.3 & \nodata &  \nodata&  \nodata &  \nodata \\
 Fe II &6383.722 &  5.553 & -2.240 & \nodata & \nodata & \nodata & \nodata & \nodata & \nodata & 36.2 & \nodata &  \nodata&  \nodata &  \nodata \\
 Fe II &6432.680 &  2.891 & -3.570 & 50.5 & \nodata & 58.6 & \nodata & \nodata & 53.6 & \nodata & 46.2 &119.2&  84.1 &  73.9 \\
 Fe II &6442.955 &  5.549 & -2.440 & \nodata & \nodata & \nodata & \nodata & \nodata & \nodata & 25.2 & \nodata &  \nodata&  \nodata &  \nodata \\
 Fe II &6456.383 &  3.903 & -2.050 & 50.1 & 39.3 & 64.7 & \nodata & \nodata & 56.4 & \nodata & 48.8 &  \nodata&  \nodata &  88.5 \\
 Fe II &6516.080 &  2.891 & -3.310 & 80.2 & \nodata & 73.5 & \nodata & \nodata & \nodata & \nodata & \nodata &  \nodata& 102.6 & 127.5 \\
 Fe II &6517.018 &  5.585 & -2.730 & \nodata & \nodata & \nodata & \nodata & \nodata & \nodata & 16.4 & \nodata &  \nodata&  \nodata &  \nodata \\
 Fe II &7222.394 &  3.889 & -3.260 & \nodata & 20.6 & \nodata & \nodata & \nodata & \nodata & 73.8 & \nodata &  \nodata&  \nodata &  \nodata \\
 Fe II &7310.216 &  3.889 & -3.370 & 28.7 & \nodata & 68.9 & \nodata & \nodata & \nodata & 66.7 & 12.4 &  \nodata&  \nodata &  \nodata \\
 Fe II &7320.654 &  3.892 & -3.230 & 96.4 & \nodata & \nodata & \nodata & \nodata & \nodata & 85.4 & \nodata &  \nodata&  \nodata &  \nodata \\
 Fe II &7449.335 &  3.889 & -3.270 & \nodata & \nodata & \nodata & \nodata & \nodata & \nodata & 68.7 & \nodata &  \nodata&  \nodata &  \nodata \\
 Fe II &7479.693 &  3.892 & -3.610 & \nodata & \nodata & \nodata & \nodata & \nodata & \nodata & 43.6 & \nodata &  \nodata&  \nodata &  \nodata \\
 Fe II &7711.724 &  3.903 & -2.500 & \nodata & \nodata & \nodata & \nodata & \nodata & \nodata & \nodata & \nodata &  \nodata&  70.6 &  71.1 \\
\enddata
\tablecomments{~ Lines listed twice correspond to those measured in adjacent orders with overlapping wavelength coverage. Log $gf$ values are taken from \citetalias{paper3} and referneces therein. }
\end{deluxetable}



\begin{deluxetable}{lcrcc}
\centering
\tablecolumns{5}
\tablewidth{0pc}
\tablecaption{Abundance Comparison \label{tab:abund_compare}}
\tablehead{ &\multicolumn{2}{c}{With C-OVER (C11a)}&
  \multicolumn{2}{c}{Without C-OVER (This Work)} \\ 
\colhead{Cluster}  &\colhead{Age (Gyrs)} &  \colhead{[Fe/H]}
&\colhead{Age (Gyrs)} & \colhead{[Fe/H]}
  }

\startdata

NGC 1718 & 1.0$-$2.5   &  $-$0.64$\pm$0.25 &1.0$-$2.5 	&$-$0.70$\pm$0.03 \\
NGC 1978 &  1.5$-$2.5  &$-$0.74$\pm$0.08  & 1.0$-$3.0           &$-$0.54$\pm$0.18\\
NGC 1866 & 0.1$-$0.3    &$+$0.04$\pm$0.04 & 0.1$-$0.5            & $-$0.27$\pm$0.20 \\
NGC 1711  & 0.06$-$0.30  &$-$0.82$\pm$0.17 &  0.06$-$0.10  & $-$0.57$\pm$0.07\\
NGC 2100  &  $<$0.04   & $<-$0.03$\pm$0.06 &   $<$0.04     &
$-$0.40$<$ [Fe/H] $<+$0.03 \\

\enddata
\tablecomments{Note that the IL abundances for NGC 1718 are shown for completeness, and that these results are not used in \textsection \ref{sec:compare} because we do not have a sample of individual stars in this cluster.}

\end{deluxetable}

\clearpage

\begin{deluxetable}{lcccccccccc}
\centering
\tablecolumns{11}
\tablewidth{0pc}
\tablecaption{Final Stellar Parameters and Fe Abundance Results \label{tab:lmcstars}}
\tablehead{
\colhead{Name}  &\colhead{$M_{V}$\tablenotemark{a}}  &\colhead{$B-V$\tablenotemark{a}} &
\colhead{$T_{eff}$} & \colhead{log(g)} & \colhead{$\xi$} &
\colhead{[Fe/H]$_{\rm{I}}$\tablenotemark{b}} & \colhead{N$_{\rm{FeI}}$}
&\colhead{[Fe/H]$_{\rm{II}}$\tablenotemark{b}}& \colhead{N$_{\rm{FeII}}$} &\colhead{v$_{r}$}\\ & & &
\colhead{(K)} & & \colhead{(kms$^{-1}$)}  &&&&& \colhead{(kms$^{-1}$)}
}
\startdata

1978-737 & $-$1.64 & 1.56 & 4079  & 0.96 & 1.70 &   $-$0.46$\pm$0.24 &69
& $-$0.43$\pm$0.34 & 17 & 290.1$\pm$0.1    \\ 

1978-730 & $-$1.86 & 1.68 &    3902  & 0.62 & 1.35 & $-$0.34$\pm$0.28
&49 & $-$0.31$\pm$0.24 & 9 & 292.9$\pm$0.1 \\ 
\\
\hline
\\

1866-954 & $-$2.64 &1.15 &4400  & 1.84 & 1.90 &     $-$0.39$\pm$0.15 &
93& $-$0.34$\pm$0.19 &24 & 300.6$\pm$0.1\\

1866-1653 & $-$3.60 &1.43 &4157  & 0.74 & 2.30 &     $-$0.31$\pm$0.19 &
26& $-$0.62$\pm$0.40 &8& 302.7$\pm$0.3\\

1866-1667 & $-$2.46 &0.96 &4802  & 1.71 & 1.95 &     $-$0.33$\pm$0.24 &
56& $-$0.47$\pm$0.41 &13 & 301.1$\pm$0.1\\
\\
\hline
\\

1711-831  &$-$3.92 & 1.40 &4050 & 0.92 & 2.15 &  $-$0.62$\pm$0.23
&59&$-$0.69$\pm$0.11& 7 & 244.1$\pm$0.1\\

1711-1194 & $-$3.97 & 1.42 & 4070 & 0.80 & 2.10 & $-$0.61$\pm$0.24&31 &
$-$0.66$\pm$0.21&6 & 247.0$\pm$0.1\\

1711-988  &  $-$4.86 & 0.50 &5850 & 1.54 & 1.75 & $-$0.46$\pm$0.11 &39&
$-$0.24$\pm$0.16&26 & 241.4$\pm$0.1\\

\\
\hline
\\

2100-b22 &   $-$5.50  & 1.51 &  3885 & 0.17 & 3.30    &
$-$0.43$\pm$0.22   & 20 &$-$0.40$\pm$0.24 & 10 & 253.5$\pm$0.1  \\

2100-c2  &  $-$6.20 &  1.38  & 4150 & 0.05 & 2.20 &  $-$0.03$\pm$0.22  &17& $-$0.07$\pm$0.20 & 3 & 242.2$\pm$0.1\\

2100-c12  & $-$5.40 &1.49 & 4000 &0.24 & 3.30 & $-$0.37$\pm$0.24&27  &
$-$0.42$\pm$0.31 &  12 & 247.2$\pm$0.1\\

\enddata
\tablenotetext{a}{V magnitudes and $B-V$ colors have been distance and reddening corrected with the ($m-M$) and  $E(B-V)$ values in Table \ref{tab:stellar_info}. }
\tablenotetext{b}{The quoted uncertainty for Fe I and Fe II abundances is the standard deviation in the abundance of all of the measured lines of each species,  $\sigma_{\rm{Fe}}$. }

\end{deluxetable}

\begin{deluxetable}{lccccc|cccc|ccc}
\centering
\tablecolumns{13}
\tablewidth{0pc}
\tablecaption{Stellar Parameter Comparison \label{tab:params}}
\tablehead{
&\multicolumn{5}{c}{Initial Photometric Parameters} &
\multicolumn{4}{c}{Spectroscopic, log$g$ constant} &\multicolumn{3}{c}{Spectroscopic, log$g$ adjusted} 
\\
\colhead{Name}  &
\colhead{$T_{\rm{phot}}$} & \colhead{log(g$_{\rm{phot}})$} & \colhead{$\xi_{\rm{Phot}}$}&
\colhead{[Fe/H]}&\colhead{[Fe/H]} 
&   \colhead{$T_{\rm{Ex}}$} 
&\colhead{$\xi_{\rm{Spec}}$}
&\colhead{[Fe/H]}&\colhead{[Fe/H]} 
&
\colhead{log(g$_{\rm{Ion}})$} &\colhead{[Fe/H]}&\colhead{[Fe/H]} \\ 
& \colhead{(K)} && \colhead{(kms$^{-1}$) } &\colhead{${\rm{I,phot}}$}
& \colhead{${\rm{II,phot}}$}
& \colhead{(K)} & \colhead{(kms$^{-1}$) }  
&\colhead{${\rm{I,spec1}}$}& \colhead{${\rm{II,spec1}}$}
&&\colhead{${\rm{I,spec2}}$}
& \colhead{${\rm{II,spec2}}$} 
}
\startdata
1978-737 & 3879 & 0.96 & 1.74 & $-$0.43 & $-$0.09 & 4079 & 1.70
&   $-$0.46 & $-$0.43
 &\nodata &   \nodata  &\nodata  \\ 

1978-730 & 3742 & 0.72 & 1.79  & $-$0.55 & $-$0.16 & 3902 & 1.35
&  $-$0.34    & $-$0.25
& 0.62& $-$0.34& $-$0.31\\ 
\\
\hline
\\

1866-954& 4469 & 1.44 & 1.64  & $-$0.29 & $-$0.54 & 4400 & 1.90
 & $-$0.45 & $-$0.58
& 1.84 &     $-$0.39&$-$0.34
\\

1866-1653 & 4057 & 0.74 & 1.79  & $-$0.08 & $-$0.32 & 4157 & 2.30
 & $-$0.31 & $-$0.62
 &\nodata &   \nodata  &\nodata 
\\

1866-1667 & 4802 & 1.71 & 1.58  & $-$0.16 & $-$0.33  & 4802 & 1.95
 & $-$0.33 & $-$0.47
&\nodata &   \nodata  &\nodata 
\\
\\
\hline
\\

1711-831& 4097 & 0.92 & 1.74  & $-$0.36 & $-$0.54 & 4050 & 2.15
& $-$0.62 & $-$0.69
&\nodata &   \nodata  &\nodata 
\\

1711-1194 & 4070 & 0.8 & 1.76  & $-$0.37 & $-$0.47 & 4070 & 2.10
& $-$0.61 & $-$0.66 
&\nodata &   \nodata  &\nodata  \\

1711-988 & 6012 & 1.54 & 1.62  & $-$0.35 & $-$0.19 & 5850 & 1.75
 & $-$0.48 & $-$0.24
&\nodata &   \nodata  &\nodata \\

\\
\hline
\\

2100-b22 & 3971 & 0.17 & 1.91  & $+$0.18 & $-$0.17 & 3885 & 3.30
 & $-$0.43 & $-$0.40
&\nodata &   \nodata  &\nodata \\

2100-c2 & 4249 & 0.05 & 1.93  & $+$0.15 & $-$0.10 & 4150 & 2.20
 & $-$0.03 & $-$0.07
&\nodata &   \nodata  &\nodata \\

2100-c12& 3997 & 0.24 & 1.89  & $+$0.05 & $+$0.03 & 4000 & 3.30
 & $-$0.37 & $-$0.42  
&\nodata & \nodata&
\nodata\\

\enddata


\end{deluxetable}


\begin{deluxetable}{lllccc}
\centering
\tablecolumns{6}
\tablewidth{0pc}
\tablecaption{Uncertainties in Fe abundance with Stellar Parameters \label{tab:uncertainties}}
\tablehead{ \colhead{Name}  & \colhead{$T_{eff}$} & \colhead{log$g$} &
  \colhead{$\xi$} & \colhead{[M/H]} \\
 & \colhead{($+150$ K)} & \colhead{($-0.5$ dex)}& \colhead{($+0.3$
   kms$^{-1}$)} &\colhead{($+0.3$ dex)}  }
\startdata

1978-737  & $+$0.02 & $-$0.10 & $-$0.11 & $+$0.07\\  
1978-730 & $-$0.09 & $-$0.18 & $-$0.10 & $+$0.00 \\

1866-954 &$+$0.04 & $-$0.07 & $-$0.12 & $+$0.05 \\
1866-1653  &$+$0.04   & $-$0.07 & $-$0.06&  $+$0.07\\ 
1866-1667 &$+$0.13 &  $+$0.02   &$-$0.10 & $+$0.01 \\

1711-831  & $-$0.02   & $-$0.10 &$-$0.12 & $+$0.07\\
1711-988  &$+$0.10  & $+$0.03     & $-$0.05   & $+$0.00\\ 
1711-1194 & $+$0.02 & $-$0.11 & $-$0.13 & $+$0.07\\ 

2100-b22 & $-$0.05 &  \nodata  & $-$0.24   &   $-$0.07           \\
2100-c2 & $-$0.05 &\nodata &$-$0.18 & $+$0.04\\
2100-c12 &$+$0.02 &$-$0.09 & $-$0.04 & $+$0.02\\ 

\enddata
\end{deluxetable}

\begin{deluxetable}{l|r|r|cr}
\centering
\tablecolumns{5}
\tablewidth{0pc}
\tablecaption{Comparison of Mean Cluster Abundances \label{tab:result}}
\tablehead{ & \colhead{ILS} & \colhead{Stars}  & \colhead{Stars} \\
\colhead{Cluster} & \colhead{This Work} & \colhead{This Work} & \colhead{Other Authors} &\colhead{Ref} }
\startdata

NGC 1978 &$-$0.54$\pm$0.18  & $-$0.40$\pm$0.08    &$-$0.38$\pm$0.12 &1\\
& & & $-$0.96$\pm$0.20 &2\\
\hline
NGC 1866  & $-$0.27$\pm$0.20 & $-$0.34$\pm$0.04    & $-$0.50$\pm$0.10&2\\
& & & $-$0.43$\pm$0.04 &3\\
\hline
NGC 1711 & $-$0.57$\pm$0.07 &   $-$0.56$\pm$0.09    & $-$0.57$\pm$0.17\tablenotemark{b}&4 \\
\hline
NGC 2100   &
$-$0.40$<$ [Fe/H] $<+$0.03 &   $-$0.40$\pm$0.04   &$-$0.32$\pm$0.03 &5\\
& & & $-$0.57$\pm$0.06 &6\\

\enddata

\tablerefs{1. \cite{2008AJ....136..375M},
  2. \cite{2000A&A...364L..19H}, 3. \cite{mucc1866},
  4. \cite{2000A&A...360..133D}, 5. \cite{1994A&A...282..717J},
  6. \cite{1999Ap&SS.265..469H}}
\tablenotetext{a}{The uncertainties listed for  the cluster metallicities in our
  work are obtained from the standard deviation in
  the individual stellar abundances from Table \ref{tab:lmcstars}.   }
\tablenotetext{b}{Note that this value is derived from Str\"{o}mgren 
  photometry, while all other stellar results from the literature are
  derived from high resolution spectra.}

\end{deluxetable}

\end{document}